\documentclass[a4paper, 12pt]{article}

\usepackage{authblk}
\usepackage[a4paper,top=2cm,bottom=2cm,left=2cm,right=2cm,marginparwidth=1.75cm]{geometry}
\usepackage[colorlinks=true, allcolors=blue]{hyperref}

\usepackage{natbib}
\usepackage[english]{babel}
\usepackage[utf8]{inputenc}
\usepackage[T1]{fontenc}
\usepackage{lmodern}

\usepackage[title]{appendix}
\usepackage{xcolor}
\usepackage{textcomp}
\usepackage{manyfoot}
\usepackage{enumitem}
\setlist[enumerate, 1]{label=(\roman*), leftmargin=*}
\usepackage{listings}
\usepackage{placeins}

\usepackage{amsmath, amssymb, amsfonts, amsthm, mathtools, nicefrac, amscd, latexsym, mathrsfs, dsfont}
\mathtoolsset{showonlyrefs}
\usepackage{MnSymbol}

\usepackage{graphicx}
\usepackage[format=hang, font=small, labelfont=bf]{caption}
\usepackage{subcaption}

\usepackage{tabularx, fix-cm, array}
\usepackage{multirow, multicol}
\usepackage{booktabs}

\usepackage[algo2e, ruled]{algorithm2e}%

\usepackage[colorlinks=true, allcolors=blue]{hyperref}
\usepackage{orcidlink}

\theoremstyle{plain}

\newtheorem{lemma}{Lemma}

\theoremstyle{remark}

\DeclareMathOperator*{\argmin}{argmin}
\DeclareMathOperator{\KL}{KL}
\newcommand*{\dd}{\mathrm{d}}
\newcommand*{\ie}{\textit{i.e.}}
\newcommand*{\eg}{\textit{e.g.}}
\newcommand*{\ess}{\mathrm{ESS}}
\newcommand*{\essmc}{\mathrm{ESS_{MC}}}

\title{Faster Hamiltonian Monte Carlo by Learning Leapfrog Scale: an offline randomized solution}
\date{}
\author[1]{Changye Wu}
\author[2,\orcidlink{0000-0003-0995-716X}]{Pierre Pudlo}
\author[1,3,\orcidlink{0000-0001-6635-3261}]{Christian P. Robert}
\author[1,4,\orcidlink{0000-0002-7813-0185}]{Julien Stoehr}

\affil[1]{\small CEREMADE, Universit\'e Paris-Dauphine, Universit\'e PSL, CNRS, 75016 Paris, France}
\affil[2]{\small Aix Marseille Univ, CNRS, I2M, 13003 Marseille, France}
\affil[3]{\small Department of Statistics, University of Warwick,  
Coventry CV4 7AL, UK}
\affil[4]{\small Universit\'e Paris-Saclay, INRAE, AgroParisTech, UMR MIA Paris-Saclay, 91120 Palaiseau, France}

\begin{document}

\maketitle

\begin{abstract}
We introduce a Hamiltonian Monte Carlo (HMC) methodology based on 
an offline empirical calibration of randomized leapfrog parameters. The approach, referred to as eHMC, where \textit{e} stands for empirical, leverages importance sampling to construct an empirical distribution on discretization parameters, thereby eliminating the need for manual burn-in diagnostics and online adaptation.
The proposal distribution used in the calibration stage is obtained via a Population Monte Carlo scheme with tempering and relies on flexible parametric variational families such as normalizing flows. 
Once the calibration stage complete, the resulting algorithm defines a homogeneous Markov chain via a mixture of HMC kernels with a fixed mixing distribution, and hence preserves the target distribution.
Numerical experiments indicate that eHMC can achieve competitive or improved sampling efficiency compared to the No-U-Turn Sampler (NUTS) in the case useful integration times can be summarized by the offline distribution. The comparison is assessed by standard efficiency metrics normalized by the number of leapfrog steps during the post-calibration sampling phase.
\end{abstract}

\section{Introduction}

Hamiltonian Monte Carlo \citep[HMC,][]{duane1987, neal2011} has emerged as an efficient Markov Chain Monte Carlo (MCMC) sampling method, which is particularly adept at dealing with high-dimensional target distributions. 
The method relies on a deterministic differential flow stemming from Hamiltonian mechanics to produce transitions across the parameter space of an augmented distribution, referred to as Hamiltonian. 
Such time-continuous dynamics leave the augmented distribution invariant; however, in practice, the latter must be discretized. 
The transitions are hence computed using a gradient-based symplectic integrator, the most popular of which is the second-order St\"ormer-Verlet also known as the leapfrog integrator \citep{hairer2003}. A Metropolis-Hastings acceptance ratio must furthermore be added towards correcting the approximate numerical solution and hence preserving the intended target measure.

Whilst the algorithm theoretically benefits from a fast exploration of the parameter space by accepting large transitions with high probability, this efficiency is marred by its high sensitivity to hand-tuned parameters.
These include the step size $\varepsilon$ of the discretization scheme, the number of steps $L$ of the integrator, and the covariance matrix $M$ of the auxiliary variables. 
Indeed, the discretization error resulting from a poorly tuned step size often leads to a low acceptance rate, since the latter is governed by the variation of the Hamiltonian between the initial and final states of the leapfrog path, which is of order $\varepsilon^2$.
Furthermore, calibrating the number of steps of the integrator is paramount to avoid slow mixing chains or to save some computation cost. 
More specifically, if the value of $L$ is not large enough, HMC may exhibit a random walk behaviour similar to that of standard Metropolis Hastings algorithms, and may thus struggle to properly explore the support of the target. 
Conversely, if the value of $L$ is too large the associated dynamic may retrace its steps back to a neighbourhood of the initial state \citep{betancourt2015}. This may result in the algorithm wasting computation efforts and diminishing mixing. Finally, the choice of a well-suited covariance matrix for the auxiliary variable can enhance both the speed and the mixing of HMC. 
One proposal in this regard is the Riemann manifold approach of \cite{girolami2011} that adapts the covariance matrix of the auxiliary variables to the curvature of the target distribution, followed by the automatic tuning strategy of \cite{wang2013}.

In terms of calibration concerns, the popular No-U-Turn Sampler \citep[NUTS,][]{hoffman2014} automatically adjusts both the step size and the number of integration steps. 
This algorithm achieves at least the same efficiency in terms of effective sample size and expected squared jump distance as compared to the standard HMC, albeit at the cost of increased computational complexity.
After selecting the step size $\varepsilon$ by primal-dual averaging \citep{nesterov2009} during a burn-in phase, NUTS dynamically adapts $L$ at each iteration 
to determine an integration length adapted to the local Hamiltonian dynamics.
This trajectory is obtained through a recursive doubling scheme that follows the Hamiltonian flow forward and backward until a no-U-turn condition is satisfied. The symmetric trajectory construction and proposal random selection yields a valid reversible transition.

Although NUTS dynamically constructs a trajectory whose length is adapted to the local geometry, the proposal is not obtained by deterministically selecting the endpoint of this trajectory. Instead, the next state is sampled from the set of candidate states generated along the trajectory using slice sampling in the original formulation \citep{hoffman2014} and multinomial sampling in the current Stan implementation. Consequently, the principal computational cost of NUTS lies in constructing the trajectory online at every iteration in order to determine an appropriate integration length.
Recent work by \citet{bourabee2025localnuts} has also investigated adaptive calibration within NUTS. They introduce local step-size adaptation steps into NUTS using a Gibbs self-tuning mechanism. Their approach performs local adaptation during the construction of the Hamiltonian trajectory while preserving the reversibility property.

The present work explores the opportunity of learning offline a distribution of integration lengths and exploiting it to define a randomized HMC scheme that recovers part of the benefits of dynamic trajectory adaptation while avoiding the systematic construction of extended trajectories during the sampling phase. Therefore, the proposal potentially reduces the online computational cost, whenever a suitable distribution of integration lengths can be learned from a preliminary, exploratory stage. However, it is not intended to provide a universal replacement for NUTS but rather stands as a mostly automated offline calibration mechanism for randomized HMC in some settings. These cover targets where a reusable distribution of integration parameters offers an adequate summary of the trajectory lengths encountered across the typical set.

The idea of randomizing elements of the Hamiltonian dynamics has already been investigated in the literature. For instance, \citet{bou2017}  propose to randomize the integration time according to a continuous distribution, yielding a Markov process with ergodicity properties and avoiding pathological periodic trajectories of standard Hamiltonian-based samplers. 
Their convergence results establish that randomization of the integration time can fundamentally alter the long-time behavior of the sampler.
While they focus on theoretical guarantees for continuous-time randomization mechanisms, our approach targets a practical,
offline, calibration of randomized discrete-time HMCs.

In this paper, we construct an empirical joint distribution on the discretization parameters $(\varepsilon,L)$. Our finite calibration phase is exploiting the U-turn criterion of NUTS on the level sets associated with a sample of points that are approximately distributed according to the target distribution. This sample is then used in a second phase to jointly simulate $(\varepsilon, L)$ at each iteration of a regular HMC scheme. 
This strategy allows us to exploit information on admissible trajectory lengths to randomize the integration parameters directly, rather than targetting the Hamiltonian flow primarily as an intermediate object for proposal selection.
Finally, as a by-product of the initial sample points and the ones obtained when visiting the level sets, we can derive a matrix $M$ adapted to the target distribution.

Unlike calibration schemes that leverage a Markovian algorithmic structure \citep[see, \eg,][]{bourabee2025localnuts, bourabee2025withinorbitadaptiveleapfrognouturn} the empirical distribution used in our method is based on independent weighted samples obtained through a Population Monte Carlo method \citep{cappe2004}, a particular instance of adaptive importance sampling. 
As a result, the calibration stage is embarrassingly parallel and can readily be distributed across CPU cores.
Furthermore, it does not require finding the scale of a burn-in period. 
The proposals used along this algorithm are adapted to align with an annealing sequence of distributions following \cite{koblents2015}.
Most crucially, the temperature schedule of this annealing sequence is itself determined automatically by relying on an effective sample size criterion, which avoids the delicate tuning of additional annealing parameters.
Our method shares a common feature with other adaptive importance sampling approaches, namely an exponentiation of the importance weights, which enables the transition from an initial proposal distribution to a distribution well fitted to the problem at hand. 
It therefore relates to annealed importance sampling \citep{neal2001}, sequential Monte Carlo samplers \citep{delmoral2006}, and more recent Entropic Mirror Descent approaches \citep{korba2022, chopin2024, cherradi2026}. 
These methods generally pursue the more ambitious objective of constructing highly efficient proposal distributions. 
In contrast, the Population Monte Carlo stage in eHMC does not aim at optimizing the proposal as accurately.
Our method only needs to produce a sufficiently representative weighted sample from which informative initial points and a pertinent empirical distribution over $(\epsilon,L)$ can be learned for the subsequent randomized HMC algorithm.

The plan of the paper is as follows. We begin with a brief description of HMC and NUTS in Section \ref{sec:HMC}. 
We then proceed to expose our method in Section \ref{sec:ehmc}. 
We first detail the construction of the joint empirical distribution on $(\varepsilon, L)$ in Section \ref{sec:calibration}, before introducing a Population Monte Carlo solution to generate the initial weighted sample paramount to our solution.
We illustrate the efficiency of the proposed algorithm compared to the NUTS version implemented in \cite{rstan} on several benchmark examples in Section \ref{sec:numeric}.

\section{Notations and conventions}

In what follows, we assume that the random variables take values in $\mathbb{R}^d$, possibly after a reparametrization, and admit a density with respect to the Lebesgue measure $\mathrm{Leb}$. We denote by $\mathbb{M}_1$ the set of probability measures on $\mathbb{R}^d$ and by $\mathbb{M}_1^+$ the subset of those that are positive almost everywhere. 
Given a probability measure $\pi\in\mathbb{M}_1$, we denote as $\mathbb{M}_{\pi}$ the set of probability measures that dominates $\pi$.
Given two measures $\mu$ and $\nu$ in $\mathbb{M}_1$, the product measure is denoted by $\mu \otimes \nu$.
We use the same notation to refer to a measure and its associated
density, meaning that if $\pi$ is absolutely continuous with respect to $\mathrm{Leb}$, $\pi(\dd x) = \pi(x) \mathrm{Leb}(\dd x)$. 
We denote $\delta_\theta$ the Dirac measure centered on a point $\theta \in\mathbb{R}^d$.

\section{Hamiltonian Monte Carlo}
\label{sec:HMC}

Consider a probability measure $\pi \in \mathbb{M}_1^+$ with continuously differentiable log-density on $\mathbb{R}^d$, and let $\widetilde{\pi}$ denote an unnormalized and tractable version of $\pi$.
HMC \citep{neal2011} generates a Markov chain on an augmented parameter space $(\theta, v)$ where $v\in\mathbb{R}^d$ is an  auxiliary momentum variable (as opposed to $\theta$ being called the position) independent of $\theta$ and distributed according to a $d-$dimensional normal distribution, $\mathcal{N}(0, M)$.
The positive definite matrix $M \in \mathcal{S}^{++}(\mathbb{R}^d)$ is referred to as the mass matrix.
We refer the reader to \cite{betancourt2018} for a discussion on the choice of other momentum distributions.
The Hamiltonian is then defined as the unnormalized negative log-version of the joint $\pi \otimes \mathcal{N}(0, M)$, namely
\begin{equation*}
H(\theta, v) = \frac{1}{2}v^{T}M^{-1}v -\log\widetilde\pi(\theta),
\end{equation*}
and the transitions of the chain derive from Hamilton's equations applied to the latter, namely
\begin{equation}
\label{eqn:hamilton-dyn}
\frac{{\rm d}\theta}{{\rm d} t} = \frac{\partial H}{\partial v}  = M^{-1}v,
\qquad
\frac{{\rm d}v}{{\rm d} t} = - \frac{\partial H}{\partial \theta} = \nabla \log\widetilde\pi(\theta).
\end{equation}
Such a mechanism aims at efficiently exploring the parameter space $\mathbb{R}^d\times\mathbb{R}^d$ as compared to standard random-walk Metropolis-Hastings proposals while yielding a marginal chain in $\theta$ with the distribution of interest.
Since the dynamic preserves the Hamiltonian, the momentum and the target density increase together and conversely, facilitating moves between low and high probability regions.

In practice, the above differential equations cannot be solved analytically and HMC samplers resort to time reversible and symplectic numerical integrators. Among these, the leapfrog integrator is commonly used for its trade-off between second order accuracy and computational cost. 
Given a discretization time-step $\varepsilon$, it yields a mapping
$F_{\varepsilon} : (\theta, v) \mapsto (\theta^\star, v^\star)$ defined by
\begin{equation*}
\begin{cases}
\theta^\star = \theta + \varepsilon M^{-1}r,\\
v^\star = r + \nicefrac{\varepsilon}{2}\nabla \log\widetilde\pi(\theta^\star),
\end{cases}
\quad\text{where}\quad
r = v + \nicefrac{\varepsilon}{2}\nabla \log\widetilde\pi(\theta).
\end{equation*}
In order to approximate the solution $\left(\theta^\star, v^\star\right)$ at time $t$, the integrator applies the mapping $L = \lfloor \nicefrac{t}{\varepsilon} \rfloor$ times. $L$ is referred to as the number of leapfrog steps. This discrete scheme serves as a proposal kernel for $(\theta, v)$ but no longer leaves the measure $\pi \otimes \mathcal{N}(0, M)$ invariant. To account for the discretization bias and to preserve the target measure, a Metropolis-Hastings correction is introduced.  A transition from $(\theta, v)$ to the proposal $\left(\theta^\star, -v^\star\right)$ 
is accepted with probability
\begin{equation}
\label{eqn:ar-HMC}
\rho\left(\theta, v, \theta^\star, v^\star\right) = 1 \wedge \exp\left\lbrace
H(\theta, v) - H\left(\theta^\star, -v^\star\right)
\right\rbrace.
\end{equation}
which implies that detailed balance is satisfied for the target $\pi \otimes \mathcal{N}(0, M)$.
The solution of \eqref{eqn:hamilton-dyn} keeps the Hamiltonian constant, meaning proposals are always accepted for the exact dynamic. After discretization, the deviation to $H(\theta, v)$ can be bounded \citep{leimkuhler2005}, 
\begin{equation*}
\left\vert  H(\theta, v) - H\left(\theta^\star, -v^\star\right) \right\vert < C\varepsilon^2.
\end{equation*}
Hence, the acceptance rate \eqref{eqn:ar-HMC} still tends to be high even for a proposal $\left(\theta^\star, -v^\star\right)$ quite far from $(\theta, v)$, provided $\varepsilon$ remains moderate. In practice, the choice of $\varepsilon$ can be done on the basis of the global acceptance rate of the sampler using the primal-dual averaging method \citep{nesterov2009}. 
The latter provides an adaptive MCMC scheme which aims at a targeted acceptance probability $p_0 \in (0,1)$. Under mild conditions, \cite{beskos2013} have shown that the optimal scaling for HMC yields an acceptance probability of around $0.65$. Though it is not necessarily a suitable choice for $p_0$ in complex models, it can serve as a useful benchmark value.

When it comes to choosing the number $L$ of integration steps, NUTS \citep{hoffman2014} is arguably the most commonly used version of HMC sampler. It eliminates the need to specify the number $L$ of integration steps by adaptively choosing the locally largest value at each iteration of the algorithm. More precisely, given a step size $\epsilon$, the current value of $\theta$ and a momentum $v$, an iteration of NUTS begins with mimicking the Hamiltonian dynamics by recursively doubling the leapfrog path, either forward or backward with equal probability, until the path begins to retrace towards the starting point.  
For the Euclidean version, this means that the backward and forward end points of the path, $(\theta^-, v^-)$ and $(\theta^+, v^+)$, satisfy 
\begin{equation}
\label{eqn:nuts-retrace}
(\theta^+ - \theta^-)^\top M^{-1}v^- <0\quad \text{  or  } \quad (\theta^+ - \theta^-)^\top M^{-1}v^+ <0. 
\end{equation}
For an extension of the aforementioned termination criterion to the non-Euclidean case, we refer the reader to \cite{betancourt2018}. The second step of NUTS consists in randomly picking one of the position-momentum pairs visited along the forward and backward leapfrog path, each pair being weighted according to the target density. In its original version, the latter was done by a slice sampling move, but most recent versions implemented in Stan \citep{stanmanual} rely on a multinomial sampling strategy. The overall process allows for detailed balance to hold and thus validates NUTS.

Compared with the standard HMC sampler, which uses a fixed length $L$ across iterations, NUTS relies on an adaptive scheme while requiring an evaluation of the Hamiltonian along its entire leapfrog path. 
Even though this leapfrog path is locally the longest possible in terms of Equation \eqref{eqn:nuts-retrace}, 
the final proposal is not obtained by deterministically taking the last point of the trajectory, but is picked at random among the states generated along the trajectory using slice-sampling in the original version or multinomial sampling in current Stan implementation.
Progressive sampling mechanisms increase the probability of selecting states from recently built subtrees,
but the distance between the proposed position and the current one may be significantly smaller than the maximal one achieved during the construction,
so part of the computational effort does not translate directly into transitions of the Markov chain.
This online construction is essential in targets where useful integration times vary strongly across the typical set, but may be unnecessary when a reusable distribution of integration times can be learned offline.

\section{Empirical Hamiltonian Monte Carlo}
\label{sec:ehmc}

Following the observation that the systematic construction of the longest admissible Hamiltonian trajectories may be unnecessary, we propose a two-stage procedure, referred to as the eHMC sampler (see Algorithm \ref{alg:ehmc}), that (i) learns a distribution over the corresponding integration parameters and (ii) reuses the latter to define a randomized version of the Hamiltonian kernel.

\begin{algorithm2e}[!ht]
\caption{eHMC sampler}
\label{alg:ehmc}
\SetKwProg{Fn}{Function}{}{}\SetKwFunction{AdaptParam}{AdaptParam}%
\KwIn{weighted sample $\{(\vartheta_1, \omega_1), \ldots, (\vartheta_N, \omega_N)\}$,
number of epochs $K\ge 1$, target acceptance probability $p_0$, initial stepsize $\varepsilon_0^+=\varepsilon_0^-=\varepsilon_0 > 0$.}

\vspace{0.5\baselineskip}
\Fn{\AdaptParam{$\vartheta$, $v$, $p_0$, $\varepsilon_0$}}{
\textbf{initialize} $\varepsilon = \varepsilon_0$\;
\Repeat{convergence}{
    \textbf{simulate} a leapfrog trajectory $(\vartheta_\ell, v_\ell)_{\ell \ge 1}$ with step size $\varepsilon$ until finding $L^\star = L(\vartheta, v, \varepsilon)$, as defined by Equation \eqref{eqn:u-turn}\;
    \textbf{update} $\varepsilon$ so that the average acceptance probability along the trajectory approaches $p_0$\;
}
\Return $(\varepsilon, L^\star, \{(\vartheta_\ell, v_\ell)\}_{1 \le \ell \le L^\star})$\;
}

\vspace{0.5\baselineskip}
\textbf{initialize}
\begin{equation}
M^{-1} = \sum_{i = 1}^N \omega_i \left(\vartheta_i - \bar{\vartheta}\right)\left(\vartheta_i - \bar{\vartheta}\right)^\top,
\qquad \bar{\vartheta} = \sum_{i = 1}^N \omega_i \vartheta_i;\;
\end{equation}

\For{$k = 1$ \KwTo $K$}{

\For{$i = 1$ \KwTo $N$}{

\textbf{sample} $v_i \sim \mathcal{N}(0,M)$\;
\textbf{compute} $(\varepsilon_i^+, L_i^+, \{(\vartheta_{i,\ell}^+, v_{i, \ell}^+)\}_{1 \le \ell \le L_i^+})$ = 
\AdaptParam($\vartheta_i$, $v_i$, $p_0$, $\varepsilon_{i-1}^+$)\;
\textbf{compute} $(\varepsilon_i^-, L_i^-, \{(\vartheta_{i,\ell}^-, v_{i, \ell}^-)\}_{1 \le \ell \le L_i^-})$ = 
\AdaptParam($\vartheta_i$, $-v_i$, $p_0$, $\varepsilon_{i-1}^-$)\;
\textbf{compute} $V_i^+$ and $S_i^+$ according to \eqref{eqn:varp-nu}, and $V_i^-$ and $S_i^-$ according to \eqref{eqn:varn-nu}\;
}
\textbf{set}
\begin{equation*}
M^{-1} = \frac{1}{\sum_{i=1}^N\omega_i (L_i^+ + L_i^-)}\sum_{i=1}^N \omega_i \left(V_i^+ + S_i^+ + V_i^- + S_i^- \right);\;
\end{equation*}
}

\vspace{0.5\baselineskip}
\textbf{initialize} $\theta^{(0)} \sim \displaystyle\sum_{i=1}^N \omega_i \delta_{\vartheta_i}$\;

\For{$t = 1$ \KwTo $T$}{
\textbf{sample} $v$ from $\mathcal{N}\left(0, M\right)$\;
\textbf{sample}
\begin{equation*}
(\varepsilon, L) \sim \widehat{\mu}_{\mathcal{T}} = \frac{1}{2}\sum_{i=1}^N \omega_i \delta_{(\varepsilon_i^+, L_i^+)}
+
\frac{1}{2}\sum_{i=1}^N \omega_i \delta_{(\varepsilon_i^-, L_i^-)};\;
\end{equation*}
\textbf{compute} $(\theta^\star, v^\star) = F_{\varepsilon}^L(\theta^{(t-1)}, v)$\;

\hspace{.5em}\textbf{set}
\begin{equation*}
(\theta^{(t)}, v^{(t)}) = 
\begin{cases}
(\theta^\star, -v^\star),&\text{with probability }1 
	\wedge \rho(\theta^{(t-1)}, v, \theta^\star, v^\star)\text{ given by \eqref{eqn:ar-HMC}},
	\\
(\theta^{(t-1)}, v),&\text{otherwise};
\end{cases}\;
\end{equation*}

}

\Return $(\theta^{(1)}, \ldots, \theta^{(T)})$\;
\end{algorithm2e}

\FloatBarrier

\subsection{Calibration stage}
\label{sec:calibration}

The first stage of eHMC relies on importance sampling \citep{kahn1949, kahn1951} to approximate the target distribution $\pi$. 
Given $N$ independent samples $(\vartheta_1, \dotsc, \vartheta_N)$ sampled from a proposal distribution $\mu\in\mathbb{M}_\pi$, the target measure $\pi$ is estimated by the empirical measure associated with the weighted sample $\{(\vartheta_1, \omega_1), \ldots, (\vartheta_N, \omega_N)\}$, where the weights are derived from the Radon--Nikodym derivative of $\pi$ with respect to the proposal distribution $\mu$, namely
\begin{equation}
\label{eqn:is-measure}
\widehat{\pi}_N = \sum_{i=1}^N\omega_i\delta_{\vartheta_i},
\qquad \omega_i \propto \frac{\widetilde\pi(\vartheta_i)}{\mu(\vartheta_i)}
\quad(i = 1,\ldots, N).
\end{equation}

\paragraph*{Empirical distribution of the integration time} The randomization of parameters $L$ and $\varepsilon$ in eHMC is based on the empirical distribution of the longest integration time associated with each position. 
Given an augmented parameter $(\theta, v)$ and a step size $\varepsilon > 0$, the maximal integration length in the direction $v$ is given by the first occurrence of a U-turn, analogous to Equation \eqref{eqn:nuts-retrace}, along the discretized Hamiltonian trajectory
\begin{equation}
\label{eqn:u-turn}
    L(\theta, v, \varepsilon) = \inf_{\ell \in \mathbb{N}}
    \left\lbrace
(\theta^\star_\ell, v^\star_\ell) = F_{\varepsilon}^\ell(\theta, v) :
(\theta^\star_\ell-\theta)^\top M^{-1}v^\star_\ell < 0
    \right\rbrace.
\end{equation}
\begin{lemma}
\label{lem:finite-turning-time-discrete}
Assume that the potential $U : \theta \mapsto -\log \widetilde\pi(\theta)$ is coercive, \ie,  $\lVert U(\theta)\rVert_2 \rightarrow \infty$ as $\lVert \theta\rVert_2 \rightarrow \infty$. Then, for all positive definite matrices $M$, there exists $\varepsilon^\star>0$, such that for all $0 < \varepsilon \le \varepsilon^\star$, $L(\theta, v, \varepsilon)$  is finite $\pi \otimes \mathcal{N}(0,M)$-almost surely.
\end{lemma}
We refer the reader to Appendix \ref{sec:finite-u-turn} for details.
Considering a set of local step sizes $\varepsilon_i \leq \varepsilon^\star$ associated with each pair $(\vartheta_i, v_i)$, $i = 1, \ldots, N$, we thus define an empirical measure on the induced set of maximal numbers of leapfrog steps
\begin{equation*}
\mathcal{L} = \{L(\vartheta_1, v_1, \varepsilon_1), \ldots, L(\vartheta_N, v_N, \varepsilon_N), L(\vartheta_1, -v_1, \varepsilon_1), \ldots, L(\vartheta_N, -v_N, \varepsilon_N)\},
\end{equation*}
namely,
\begin{equation*}
    \widehat{\mu}_{\mathcal{L}} = \frac{1}{2}\sum_{i = 1}^N \omega_i \delta_{L(\vartheta_i, v_i, \varepsilon_i)} 
    + \frac{1}{2}\sum_{i = 1}^N \omega_i \delta_{L(\vartheta_i, -v_i, \varepsilon_i)}.
\end{equation*}

However, our aim is to further select the step size at random within the HMC scheme. 
We thus build an empirical distribution that accounts for a local adaptation of the step size for each realisation $\vartheta_i$.

Consider an excursion on the discretized level set starting from a point $(\theta, v)$. 
The leapfrog integrator generates a sequence of pairs $(\theta_{\ell}, v_{\ell}) = F_{\varepsilon}^\ell(\theta, v)$, $\ell \in\mathbb{N}^*$. 
Each of these defines a valid Metropolis--Hastings proposal for the augmented target distribution based on the acceptance probability $\rho(\theta, v, \theta_{\ell}, v_{\ell})$ as defined in Equation \eqref{eqn:ar-HMC}. 
Hence, we can locally adapt the step size by monitoring the average acceptance probability along the leapfrog path.
Namely, given a user-specified target probability $p_0 \in (0, 1)$, we calibrate the step size from
\begin{equation}
\label{eqn:updt-eps}
    \overline{\varepsilon}(\theta, v, p_0) = \sup\left\lbrace 
    0 < \varepsilon \le \varepsilon^\star : 
    \frac{1}{L(\theta, v, \varepsilon)} 
    \sum_{\ell = 1}^{L(\theta, v, \varepsilon)}
     \rho\left(\theta, v, \theta_{\ell}, v_{\ell}\right)
 \ge p_0
    \right\rbrace.
\end{equation}
The probability $p_0$ plays a role similar to target acceptance probability in the primal-dual averaging algorithm.
The latter set is non empty. Indeed, for any time interval, the deviation to $H(\theta, v)$ is of order $\mathcal{O}(\varepsilon^2)$ \citep{leimkuhler2005}, which ensures that for $\varepsilon$ small enough the acceptance probability \eqref{eqn:ar-HMC} becomes larger than $p_0$.
Setting $\varepsilon_i^+ = \overline{\varepsilon}(\vartheta_i, v_i, p_0)$ and $\varepsilon_i^- = \overline{\varepsilon}(\vartheta_i, -v_i, p_0)$, the empirical measure on the discretization parameters is then defined as
\begin{equation}
\label{eqn:dist-time}
    \widehat{\mu}_{\mathcal{T}} = \frac{1}{2}\sum_{i = 1}^N \omega_i \delta_{(\varepsilon_i^+, L(\vartheta_i, v_i, \varepsilon_i^+))} + \frac{1}{2}\sum_{i = 1}^N \omega_i \delta_{(\varepsilon_i^-, L(\vartheta_i, -v_i, \varepsilon_i^-))}.
\end{equation}

The U-turn criterion in \eqref{eqn:nuts-retrace} is chosen here as a practical device for constructing the empirical distribution \eqref{eqn:dist-time}. Its validity is theoretically motivated by Lemma \ref{lem:finite-turning-time-discrete}, which establishes that the corresponding stopping time is almost surely finite and therefore induces a well-defined empirical distribution over $(\varepsilon,L)$. Note that the U-turn criterion is not essential to the proposed methodology. Any alternative offline stopping criterion could be used, provided an analogous result holds, namely that this criterion defines an almost surely finite stopping time and hence induces a well-defined distribution over $(\varepsilon,L)$. Once such a distribution is obtained, the validity of the final sampler no longer depends on the particular criterion at its core (see Section \ref{sec:sampling-stage} for the stationarity argument). Alternative criteria satisfying the above requirement may then prove useful, especially in settings where the first U-turn time is not the most informative proxy for a suitable integration length, such as highly oscillatory Hamiltonian dynamics. They may further bring an assessment of convergence by comparing the resulting outcomes.

\paragraph*{Adapting the mass matrix} The Hamiltonian dynamic and consequently the empirical distribution $\widehat{\mu}_{\mathcal{T}}$ depend on the choice of the mass matrix $M$.
It is well known that tuning $M$ so that $M^{-1}$ accounts for the covariance structure of the target distribution $\pi$ can substantially improve the efficiency of the sampler.
Accordingly, we first estimate $M^{-1}$ by the sample covariance of the weighted particles $\{(\vartheta_1, \omega_1), \ldots, (\vartheta_N, \omega_N)\}$, namely
\begin{equation*}
M^{-1} = \sum_{i = 1}^N \omega_i \left(\vartheta_i - \bar{\vartheta}\right)\left(\vartheta_i - \bar{\vartheta}\right)^\top,
\qquad
\bar{\vartheta} = \sum_{i = 1}^N \omega_i \vartheta_i,
\end{equation*}
which corresponds to the minimizer of the Kullback--Leibler divergence over Gaussian
momentum distributions \citep{girolami2011}.
The latter can then be refined by exploiting the additional information provided by all the positions traced out by the integrator during the construction of the empirical measure \eqref{eqn:dist-time}. 
Indeed, for each initial pair $(\vartheta_i, v_i)$, consider the sequences of position-momentum  $(\vartheta_{i,\ell}^+, v_{i,\ell}^+) = F_{\varepsilon_i^+}^\ell(\vartheta_i, v_i)$ and  $(\vartheta_{i,\ell}^-, v_{i,\ell}^-) = F_{\varepsilon_i^-}^\ell(\vartheta_i, -v_i)$ generated along the leapfrog trajectory. 
In the continuous-time, ideal, setting, these intermediate positions are distributed according to the same marginal distribution as $\vartheta_i$. 
In the discretized setting, we account for integration error by weighting each position according to its corresponding Metropolis--Hastings acceptance probability \eqref{eqn:ar-HMC}.
The discrete measure
\begin{multline}
\label{eqn:emp-measure-mass-matrix}
    \nu \propto \sum_{i = 1}^N \omega_i \Bigg[
    \sum_{\ell = 1}^{L(\vartheta_i, v_i, \varepsilon_i^+)} 
    \left\lbrace
    \delta_{\vartheta_{i, \ell - 1}^+} + \rho(\vartheta_i, v_i, \vartheta_{i,\ell}^+, v_{i, \ell}^+)\left(\delta_{\vartheta_{i, \ell}^+} - \delta_{\vartheta_{i, \ell - 1}^+}\right)
    \right\rbrace
    \\
    + \sum_{\ell = 1}^{L(\vartheta_i, -v_i, \varepsilon_i^-)} 
    \left\lbrace
    \delta_{\vartheta_{i, \ell - 1}^-} + \rho(\vartheta_i, -v_i, \vartheta_{i,\ell}^-, v_{i, \ell}^-)\left(\delta_{\vartheta_{i, \ell}^-} - \delta_{\vartheta_{i, \ell - 1}^-}\right)
    \right\rbrace
    \Bigg].
\end{multline}
yields a Rao–Blackwellized empirical covariance associated with the Metropolis–Hastings transitions along each trajectory that we used to estimate $M^{-1}$ (see Appendix \ref{sec:formula}).

The calibration scheme thus consists of a series of adaptation epochs that alternate between updating first the mass matrix and then constructing the empirical measure \eqref{eqn:dist-time} of the discretization parameters. 
For the sake of simplicity, we have described the Rao--Blackwellization based on the positions of the last excursion of each trajectory. 
In practice, the computation of $\varepsilon_i$, as defined in Equation \eqref{eqn:updt-eps} typically involves multiple excursions on the same level set, and all intermediate positions can be incorporated in the estimator in the same manner.

\FloatBarrier

\subsection{Sampling stage}
\label{sec:sampling-stage}

In standard MCMC methods, a preliminary burn-in phase is typically required to ensure that the chain has moved sufficiently far from its initial distribution toward its invariant distribution. This transient regime is often difficult to diagnose and may incur a non-negligible computational cost. 
In standard NUTS implementations, this warmup phase is used to tune algorithmic parameters such as the step size and mass matrix.
In eHMC, this role is transferred to the offline calibration stage. The latter provides an importance-weighted sample that yields an empirical approximation of $\pi$, removing the reliance on a transient Markov-chain phase for parameter adaptation and initialization.

Once both $\widehat{\mu}_T$ and $M$ are set, they are used to define a homogeneous Markov chain.
At each iteration, say $k$, a pair $(\varepsilon_k, L_k)$ is sampled from $\widehat{\mu}_{\mathcal{T}}$ and an HMC proposal is generated using the leapfrog integrator with mass matrix $M$ and integration length $L_k\varepsilon_k$ (see Algorithm \ref{alg:ehmc}). 
The resulting algorithm yields a randomized HMC kernel that does not belong to the class of adaptive MCMC methods with online parameter updates. Instead, it defines a mixture of Markov transition kernels with fixed mixing distribution $\widehat{\mu}_T$.
Since, for any fixed $(\varepsilon, L)$, the corresponding HMC kernel leaves the target distribution $\pi$ invariant \citep{neal2011}, it follows that the mixture kernel defined by sampling $(\varepsilon, L)$ from $\widehat{\mu}_{\mathcal{T}}$ also preserves $\pi$ \citep{tierney1994, roberts2007}.

\subsection{Construction of the importance sampling proposal}
\label{sec:pmc}

The eHMC sampler relies on having an initial proposal distribution $\mu$ to derive a weighted sample approximately distributed according to $\pi$. 
Finding a well suited proposal $\mu$ is paramount to guarantee good statistical properties of importance sampling estimators \citep[\eg,][]{agapiou2017, chatterjee2018}, but this may prove challenging.
Numerous adaptive strategies have been proposed over the past two decades to design efficient proposal distributions, and we refer the reader to \cite{elvira2021} for a broad overview.
However the objective here differs from standard importance sampling. 
Rather than seeking highly accurate approximations of the target distribution in order to control bias and variance, our goal is to obtain a weighted sample that captures sufficient diversity in regions of non-negligible probability mass.

In what follows, we rely on a Population Monte Carlo (PMC) scheme \citep{cappe2004} to construct the proposal $\mu$ (see Algorithm \ref{alg:pmc}).
PMC is an adaptive importance sampling framework that generates a sequence $(\mu_t)_{t \ge 0}$ in $\mathbb{M}_\pi$ by iteratively updating the proposal $\mu_{t+1}$ using weighted samples drawn from $\mu_t$. 
We adopt a tempering strategy inspired by \cite{koblents2015} and \cite{aufort2022}, which improves robustness in complex or multimodal settings.
The PMC procedure described below should be viewed as one possible solution for the offline calibration stage rather than as an essential component of the proposed methodology. More generally, any procedure yielding a representative weighted sample of the target distribution could be used to estimate the empirical distribution \eqref{eqn:dist-time}. The purpose of the calibration stage is to transfer the adaptation effort outside the final Markov chain, not to advocate a particular adaptive importance sampling method.
This does not, however, eliminate all calibration choices: the quality of the initial proposal, the PMC settings (or those of any alternative calibration procedure), and the size of the calibration sample remain important practical considerations, as it is generally the case for adaptive importance sampling methods.

\paragraph*{Proposal adaptation} The initial proposal $\mu_0$ is set by the user. 
In a Bayesian context, a natural choice is the prior distribution. 
Subsequent proposals are chosen from a parametric family $\mathcal{F} \subset \mathbb{M}_\pi$ from which we can sample and evaluate the density.
To progressively bridge the gap between $\mu_t$ and $\pi$, we introduce a sequence $(\pi_t)_{t \ge 1}$ of annealed target distributions defined, for all $t\ge 0$, as
\begin{equation}
\label{eqn:pi-t}
    \pi_{t+1} \propto \widetilde\pi^{\beta_{t+1}} \mu_{t}^{1-\beta_{t+1}},
\end{equation}
where $(\beta_t)_{t \ge 1} \in[0,1]^{\mathbb{N}}$ is a non-decreasing sequence such that $\beta_1 = 0$ and $\lim_{t \to \infty} \beta_t = 1$. 
This construction defines a continuum of intermediate distributions connecting $\mu_0$ to $\pi$ that has been extensively used in the tempering literature \citep{chopin:papaspiliopoulos:2021} and helps to uncover regions of $\mathbb{R}^d$ where $\pi$ has non-negligible mass. 
At iteration $t\ge 0$, given a sample $(\vartheta_{t,1}, \ldots, \vartheta_{t,N_0})$ from $\mu_t$, the annealed target $\pi_{t+1}$ is approximated by the self-normalized importance sampling measure
\begin{equation}
\label{eqn:is-measure-temp}
    \widehat{\pi}_{t+1} \propto \sum_{i=1}^{N_0} \widetilde{\omega}_{t,i}^{\beta_{t+1}} \delta_{\vartheta_{t,i}},
    \qquad 
    \widetilde{\omega}_{t,i} = \frac{\widetilde{\pi}(\vartheta_{t,i})}{\mu_t(\vartheta_{t,i})}.
\end{equation}
The next proposal is then obtained as the projection of $\widehat{\pi}_{t+1}$ onto $\mathcal{F}$
\begin{equation}
\label{eqn:updt-mu}
    \mu_{t+1} = \argmin_{\mu \in \mathcal{F}} \KL(\widehat{\pi}_{t+1} \parallel \mu)
    = \argmin_{\mu \in \mathcal{F}} \left\lbrace - \sum_{i = 1}^{N_0} \widetilde{\omega}_{t,i}^{\beta_{t+1}} \log \mu(\vartheta_{t, i})\right\rbrace.
\end{equation}
When $\mathcal{F}$ is parametric, this corresponds to computing a maximum likelihood estimate of the family parameter based on the weighted sample. 

\paragraph*{Temperature scheduling} A standard strategy to define the sequence $(\beta_t)_{t\ge 1}$ consists in selecting $\beta_{t+1}$ to control the effective sample size (ESS) of the importance weights \citep{gramacy2010, beskos2016, elvira2021}. 
Assume that at iteration $t$, for all $\beta\in[0,1]$, $\widetilde\pi/\mu_t \in L^{2\beta}(\mu_t)$.
Then, the ESS associated with proposal $\mu_t$ and unnormalized target $\mu_{t}^{1-\beta} \widetilde\pi^{\beta}$ is 
\begin{equation*}
    \ess\left( \widetilde\pi^{\beta} \mu_{t}^{1-\beta}, \mu_t\right) = N \frac{\left\lbrace\int_{\mathbb{R}^d} \widetilde{\omega}_t(\vartheta)^\beta\mu_t(\dd \vartheta)\right\rbrace^2}{\int_{\mathbb{R}^d} \widetilde{\omega}_t(\vartheta)^{2\beta}\mu_t(\dd \vartheta)},
    \qquad\widetilde{\omega}_t(\cdot) = \frac{\widetilde\pi(\cdot)}{\mu_t(\cdot)}.
\end{equation*}
Its estimate
\begin{equation}
\label{eqn:ess-is}
    \widehat{\ess}\left(\widetilde\pi^{\beta} \mu_{t}^{1-\beta} , \mu_t\right) = \frac{\left(\sum_{i=1}^{N_0} \widetilde{\omega}_{t,i}^\beta \right)^2}{\sum_{i=1}^{N_0} \widetilde{\omega}_{t,i}^{2\beta}},
\end{equation}
quantifies the discrepancy between the weighted and unweighted empirical measures, or, equivalently, the degeneracy of the importance weights: a higher effective sample size indicates a better empirical approximation of the target distribution.
In our setting, $\beta_{t+1}$ is chosen to ensure that the reweighted sample provides a sufficiently accurate approximation of the annealed target $\pi_{t+1}$, and avoids overfitting to the current sample. 
More precisely, we set
\begin{equation}
\label{eqn:updt-beta}
    \beta_{t+1} = \sup \left\{
    \beta \in [0,1] : 
    \widehat{\ess}\left(\widetilde{\pi}^\beta \mu_t^{1-\beta}, \mu_t\right) \ge \lambda_0 N_0
    \right\},
\end{equation}
where $\lambda_0 \in (0, 1)$ is a user-specified parameter.
Under the assumption $\widetilde\pi/\mu_t \in L^{2\beta}(\mu_t)$, the function  $\beta \mapsto \ess(\pi^\beta \mu_t^{1-\beta}, \mu_t)$ is continuous and non-increasing (see Appendix \ref{sec:monotonicity}), which ensures that this update is well-defined. 
The adaptation stops when we obtain $\widehat{\ess}\left(\widetilde{\pi}, \mu_t\right) \ge \Delta$, where $\Delta \leq \lambda_0 N_0$ is a user-specified threshold.

\begin{algorithm2e}[!t]
\caption{Population Monte Carlo initializer}
\label{alg:pmc}
\SetKwProg{Fn}{Function}{}{}\SetKwFunction{ImportanceSampling}{ImportanceSampling}%
\SetKwProg{Fn}{Function}{}{}\SetKwFunction{ResampleStep}{ResampleStep}%
\KwIn{Initial proposal $\mu_0 \in \mathbb{M}_\pi$, proposal family $\mathcal{F}$, number of particles $N$, number of training particles $N_0$, ESS threshold $\Delta$, Markov kernels $(P_t)_{t\ge 1}$ preserving $(\pi_t)_{t\ge 1}$.}

\vspace{0.5\baselineskip}
\Fn{\ImportanceSampling{$N$, $\mu$}}{
\textbf{sample} $(\vartheta_{1}, \ldots, \vartheta_{N}) \sim \mu$\;
\textbf{compute} 
$\displaystyle{\widetilde{\omega}_{i} = \frac{\widetilde\pi(\vartheta_{i})}{\mu(\vartheta_{i})},
\quad(i = 1, \ldots, N),\quad\text{and}\quad
\widehat{\ess}(\widetilde\pi, \mu) = \frac{\left( \sum_{i = 1}^N \widetilde{\omega}_i \right)^2}{\sum_{i = 1}^N \widetilde{\omega}_i^2}}$\;

\Return $\{(\vartheta_{1}, \widetilde\omega_1), \ldots, (\vartheta_{N}, \widetilde\omega_{N}), \widehat{\ess}(\widetilde\pi, \mu)\}$\;
}
	
\vspace{0.75\baselineskip}
\Fn{\ResampleStep{$(\vartheta_{1}, \omega_1), \ldots, (\vartheta_N, \omega_N)$, $P$}}{
\textbf{sample} 
$\displaystyle{\bar{\vartheta}_{i} \sim 
\frac{1}{\sum_{i = 1}^{N} \omega_i}\sum_{i=1}^{N}\omega_i\delta_{\vartheta_{i}},
\quad(i=1,\ldots, N)}$\;

\If{Markov perturbation is used}{
\textbf{sample} $(\bar{\vartheta}_{1}',\ldots,\bar{\vartheta}_{N}') \sim P(\bar{\vartheta}_{1},\cdot) \otimes \ldots \otimes P(\bar{\vartheta}_{N},\cdot)$\;
\textbf{set} $(\bar{\vartheta}_{1},\ldots,\bar{\vartheta}_{N}) = (\bar{\vartheta}_{1}',\ldots,\bar{\vartheta}_{N}')$\;
}
\Return $(\bar{\vartheta}_{1},\ldots,\bar{\vartheta}_{N})$\;
}

\vspace{\baselineskip}
\textbf{initialize} $t = 0$\;
\textbf{compute} 
$\{(\vartheta_{1}, \widetilde\omega_1), \ldots, (\vartheta_{N_0}, \widetilde\omega_{N_0}), \widehat{\ess}(\widetilde\pi, \mu_t)\} = \ImportanceSampling(N_0, \mu_t)$\;

\Repeat{$\widehat{\ess}(\widetilde\pi, \mu_t) \ge \Delta$}{
\textbf{increment} $t = t + 1$\;
\textbf{update} the inverse temperature $\beta_t$ according to \eqref{eqn:updt-beta}\;
    
\If{resampling is used}{
\textbf{sample} $(\bar{\vartheta}_{1},\ldots,\bar{\vartheta}_{N_0}) = 
\ResampleStep((\vartheta_{1}, \widetilde\omega_1^{\beta_{t}}), \ldots, (\vartheta_N, \widetilde\omega_{N_0}^{\beta_{t}}), P_t)$\;

\textbf{set} $(\vartheta_{1}, \ldots, \vartheta_{N_0}) = (\bar{\vartheta}_{1},\ldots,\bar{\vartheta}_{N_0})$ and $(\widetilde\omega_1, \ldots, \widetilde\omega_{N_0}) = (1, \ldots, 1)$\;
}
\textbf{update} the proposal distribution
$\displaystyle{\mu_{t} = \argmin_{\mu \in \mathcal{F}} \left\lbrace
-\sum_{i = 1}^{N_0} \widetilde{\omega}_i^{\beta_t} \log \mu(\vartheta_i)
\right\rbrace}$\;
\textbf{compute} $\{(\vartheta_{1}, \widetilde\omega_1), \ldots, (\vartheta_{N_0}, \widetilde\omega_{N_0}), \widehat{\ess}(\widetilde\pi, \mu_t)\} = \ImportanceSampling(N_0, \mu_t)$\;
}

\vspace{.5\baselineskip}
\textbf{compute} $(\vartheta_{1}, \widetilde\omega_1), \ldots, (\vartheta_{N}, \widetilde\omega_{N}) = \ImportanceSampling(N, \mu_t)$\;
\uIf{resampling is used}{
\textbf{sample} $(\bar{\vartheta}_{1},\ldots,\bar{\vartheta}_{N}) = 
\ResampleStep((\vartheta_{1}, \widetilde\omega_1), \ldots, (\vartheta_N, \widetilde\omega_{N}), P_t)$\;
\textbf{set} $(\vartheta_{1}, \ldots, \vartheta_{N}) = (\bar{\vartheta}_{1},\ldots,\bar{\vartheta}_{N})$ and $(\widetilde\omega_1, \ldots, \widetilde\omega_{N}) = (1, \ldots, 1)$\;
}
\textbf{set} $\omega_i = \widetilde\omega_i \Big/ \sum_{j=1}^{N} \widetilde\omega_j$, $(i = 1, \ldots, N)$\;

\Return $\left\lbrace (\vartheta_1, \omega_1),\ldots,(\vartheta_{N}, \omega_{N})\right\rbrace$\;
\end{algorithm2e}

\paragraph*{Practical implementation} 
Unlike standard adaptive importance sampling methods, where the effective sample size is controlled to ensure that a large proportion of particles carry significant weight, our objective is more modest. The threshold $\Delta$ is chosen only to guarantee a minimal level of effectiveness, sufficient to capture regions of non-negligible probability mass. The importance weights may then exhibit substantial variability.
To mitigate the resulting weight degeneracy, we can perform a resampling step from the empirical measure $\widehat{\pi}_{t+1}$,
possibly followed by a Markov kernel perturbation that preserves $\pi_{t+1}$ in order to prevent sample impoverishment (see Algorithm \ref{alg:pmc}). 
This resample-move strategy, commonly used in sequential Monte Carlo methods \citep{chopin:papaspiliopoulos:2021}, produces a set of unweighted particles while preserving consistency with respect to the target $\pi_{t+1}$. 
Consequently, the proposal update step \eqref{eqn:updt-mu} consists in fitting $\mu_{t+1}$ to the sample that has undergone this transformation, rather than to the original weighted sample.

In practice, this procedure can also be applied at the final iteration of the adaptation, that is once we get $\mu$ such that  $\widehat{\ess}\left(\widetilde{\pi}, \mu\right) \ge \Delta$, so that the sample used in the calibration phase consists of independent and equally weighted particles, namely 
\begin{equation*}
\widehat{\pi}_N = \frac{1}{N} \sum_{i=1}^N \delta_{\vartheta_i}.
\end{equation*}

\section{Numerical experiments}
\label{sec:numeric}

In this section, we assess the performance of the proposed eHMC algorithm, in comparison to NUTS, on four target distributions chosen to illustrate different challenges, including strong nonlinearity, correlation structure and multiscale geometry. All experiments were implemented using a combination of \textsf{R} and \textsf{Python}, with Stan models used to define the target distributions. We used the version of NUTS implemented in \texttt{Rstan} \citep{rstan}. The codes for eHMC and the experiments are available on \url{https://github.com/jstoehr/eHMC}. 

\subsection{Models}

\paragraph*{Banana-shaped distribution (\textsf{banana})} We first consider a two-dimensional nonlinear target distribution defined by
\begin{equation*}
\theta_1 \sim \mathcal{N}(0, \sigma_1^2), \qquad
\theta_2 \mid \theta_1 \sim \mathcal{N}\left(\lambda(\theta_1^2 - \ell), \sigma_2^2\right),
\end{equation*}
with $\sigma_1 = 10$, $\lambda = 0.03$, $\ell = 100$ and $\sigma_2 = 1$. 
This distribution is commonly used as a benchmark for testing the robustness of sampling algorithms to nonlinear dependencies \citep{naderi:etal:2026}.

Within the eHMC calibration stage, the initial proposal is set to a multivariate Gaussian distribution with mean $(0, -\lambda\ell)^{\top}$ and diagonal covariance matrix $\sigma_1^2 I_2$. The latter centers the proposal along the main curvature of the target while maintaining isotropic variability.

\paragraph*{Multivariate Normal distribution (\textsf{MVNorm})} We next consider a $100$-dimensional Gaussian target distribution with density
\begin{equation*}
\pi(\theta) \propto \exp\left(-\frac{1}{2} \theta^\top A^{-1} \theta \right),
\qquad A_{i,j} = 0.99^{|i-j|},\quad (i,j=1,\cdots, 100).
\end{equation*}
The covariance matrix is taken as the AR(1) correlation matrix with autoregressive parameter $\rho = 0.99$.
The resulting matrix exhibits a condition number of approximately $7.6 \times 10^3$, indicating a moderately ill-conditioned geometry with strong anisotropy due to strong dependence between neighboring coordinates.

The initial proposal for eHMC is set to a mixture of two Gaussian distributions centered at zero,
\begin{equation*}
\mu_0 = \alpha \mathcal{N}(0, \Sigma_1) + (1-\alpha) \mathcal{N}(0, \Sigma_2),
\qquad \alpha = 0.1,
\end{equation*}
where the covariance matrices $\Sigma_1$ and $\Sigma_2$ are constructed from the eigendecomposition $A = Q D Q^\top$ of the target covariance matrix. Specifically, $\Sigma_1$ and $\Sigma_2$ are defined as
\begin{equation*}
\Sigma_1 = Q \operatorname{diag}(d_1, \ldots, d_{100}) Q^\top,
\qquad
\Sigma_2 = Q \operatorname{diag}(\lambda_1,\ldots, \lambda_{100}) Q^\top,
\end{equation*}
with for $j = 1, \ldots, 100$
\begin{equation*}
d_j =
\begin{cases}
\log(D_{jj}), & D_{jj} > 1,\\
D_{jj}, & D_{jj} \le 1,
\end{cases}
\qquad
\lambda_j =
\begin{cases}
0.6 D_{jj}, & D_{jj} > 1,\\
1.1 D_{jj}, & D_{jj} \le 1.
\end{cases}
\end{equation*}
This construction yields a dispersed proposal that captures different scales of the target distribution.

\paragraph*{Bayesian logistic regression (\textsf{BLP})} We also consider a Bayesian logistic regression model with
\begin{equation*}
\theta \sim \mathcal{N}(0, I),
\qquad
y_i \sim \mathrm{Bernoulli}\left(\mathrm{logit}^{-1}(x_i^\top \theta)\right),
\qquad i=1,\ldots,n,
\end{equation*}
This model yields a posterior distribution that is log-concave but may exhibit strong correlations between parameters depending on the design matrix.
The inference is based on the German Credit dataset from the UCI Machine Learning Repository \citep{dua2019}, which consists of $n = 1000$ observations. The design matrix includes an intercept term and $24$ covariates. 

\FloatBarrier

The initial proposal distribution is constructed as a Gaussian approximation centered around a ridge-regularized least-squares estimate of the posterior mode. Specifically, we define the approximate precision matrix as $A = X^\top X + \sigma^{-2} I_K$.
The proposal mean is then given by $A^{-1} X^\top y$,
with covariance matrix $cA^{-1}$, where $c > 1$ is an inflation factor introduced to broaden the proposal and improve coverage of the posterior tails. In what follows, we set $c=4$.
This construction provides a computationally inexpensive Laplace-style initialization that is substantially better adapted to the posterior than an isotropic prior proposal.

\paragraph*{Neal's funnel distribution (\textsf{funnel})}
To assess limitations of the eHMC solution, we consider Neal's funnel distribution. For
dimension $d = 10$, the target is defined by
\begin{equation*}
  y \sim \mathcal{N}(0,3^2), \qquad
  x_i \mid y \sim \mathcal{N}(0,\exp(y)), \quad i=1,\ldots,d-1 .
\end{equation*}
Equivalently, up to an additive constant,
\begin{equation*}
  \log \pi(y,x)
  \propto
  -\frac{1}{2}\frac{y^2}{9}
  -\frac{d-1}{2}y
  -\frac{1}{2}\exp(-y)\sum_{i=1}^{d-1}x_i^2 .
\end{equation*}
This target is a standard challenge for randomized HMC methods since the curvature varies strongly with the value of $y$. For some parts of the distribution, small changes in $x$ require small integration time, whereas for other parts much larger moves are possible. We therefore expect dynamic trajectory construction to be more important on this target than on the previous benchmarks.

The initial proposal distribution uses the correct conditional structure but inflated conditional variance,
\begin{equation*}
  y \sim \mathcal{N}(0,3^2), \qquad
  x_i \mid y \sim \mathcal{N}(0,4\exp(y)).
\end{equation*}
Although this proposal captures the characteristic funnel geometry, it remains a poor approximation of the target distribution. In particular, the variance inflation produces a substantial mismatch with the target, resulting in an effective sample size close to one particle in some experiments.).

\subsection{Experimental design}

For each model, the experiments were repeated over $20$ seeds. 
For each seed, we ran $50$ parallel chains and set the target acceptance probability $p_0 \in \{0.651, 0.80\}$ for both methods.

\paragraph*{NUTS setting} We use the NUTS implementation in Stan. Following the default choice, we used a diagonal mass matrix $M$. Each chain then consists of $4000$ iterations in total, including $2000$ warmup iterations during which we used the default dual averaging adaptation for the step size, and $T = 2000$ kept samples.
The computational cost of NUTS is measured by the number of gradient evaluations, or equivalently, up to a constant factor, by the total number of leapfrog steps performed during the sampling phase. The warmup phase is not included in this cost.

\paragraph*{eHMC setting} For each seed, the calibration phase of eHMC was performed using the PMC procedure described in Section \ref{sec:pmc} with an initial sample size $N_0 = 50000$, $\lambda_0 = 0.8$ and an effective sample size threshold $\Delta \in \{0.05 N_0, 0.1 N_0\}$. 
The parametric family $\mathcal{F}$ was defined through a normalizing flow, which provides a flexible class of tractable distributions based on an invertible and differentiable transformation that maps a simple base distribution to the distribution of $(\vartheta_{t,1}, \ldots, \vartheta_{t, N_0})$. 
More precisely, let $\{f_\xi : \mathbb{R}^d \to \mathbb{R}^d\}_{\eta \in \Xi}$ be a family of bijective and differentiable mappings and $p$ be a base distribution. The family $\mathcal{F}$ is defined as
\begin{equation*}
\mathcal{F} = \left\{\theta\mapsto p\left(f_\xi^{-1}(\theta)\right) \left\lvert \det D f_\xi^{-1}(\theta) \right\rvert\;;\; \xi\in\Xi\right\},
\end{equation*}
where $D f_\xi^{-1}$ is the Jacobian of $f_\xi^{-1}$. 
In our setting, the base distribution was set to a multivariate Gaussian $\mathcal{N}(0, \Sigma_t)$, where $\Sigma_t$ denotes the covariance matrix associated with the empirical measure \eqref{eqn:is-measure-temp}. 
The normalizing flow was implemented as a Masked Autoregressive Flow \citep[MAF,][]{papamakarios2017} with $5$ transformations, each parameterized by neural networks with $2$ hidden layers of $32$ units.
The model was trained using mini-batches of size $2048$ and a learning rate of $10^{-3}$. Early stopping was used with a patience of $20$ epochs without validation improvement, reserving $20\%$ of the data for validation. 
The procedure was implemented using \texttt{zuko v1.4.0} \citep{rozet2022zuko}.

We stress anew that the fitted normalizing flow is not intended to serve as a transport map for running HMC on a transformed target density. Within the PMC tempering scheme, the flow is successively fitted to a sequence of annealed target distributions to improve the efficiency of the importance sampler. However, the calibration procedure stops as soon as the proposal yields a sufficiently large importance sampling ESS with respect to the target $\pi$. Consequently, the annealing sequence does not need to reach $\beta=1$, and the fitted flow approximates the last annealed distribution visited by the algorithm rather than the target distribution $\pi$.

After the proposal adaptation step, we resampled $N\in\{2000, 5000\}$ particles, without applying a Markov kernel perturbation. 
This equally weighted sample was then used to construct the empirical distribution $\widehat{\mu}_{\mathcal{T}}$, as defined in \eqref{eqn:dist-time}, with $K = 1$ adaptation epoch.
To match the default choice in \texttt{Rstan}, we used a diagonal mass matrix obtained from the diagonal of the empirical covariance estimate based on the empirical measure \eqref{eqn:emp-measure-mass-matrix}.
The final eHMC chains are then run for $T=2000$ iterations without further online adaptation.
The computational cost of eHMC is measured by the number of gradient evaluations, or equivalently, up to a constant factor, by the total number of leapfrog steps performed during the sampling phase. The offline calibration stage is not included in this cost.

\subsection{Performance metrics}

For each model, we compare eHMC and NUTS in terms of the effective sample size associated with the Markov chain and the expected squared jump distance.
To account for differences in computational cost across samplers, performances are assessed by normalizing both metrics by 
the number of gradient evaluations of the log-target density, or equivalently the total number of leapfrog steps performed.
This normalization reflects the fact that the evaluation of $\nabla \log\widetilde\pi$ generally dominates the overall computational cost.

\paragraph*{Markov chain effective sample size ($\essmc$)} We recall that the standard effective sample size associated with a Markov chain $\{\theta^{(1)}, \ldots, \theta^{(T)}\}$ and a measurable function $f$ is defined as
\begin{equation*}
\essmc(f) = {T}\Bigg/{\left(1+2\sum_{k=1}^{\infty}\rho_k(f)\right)},
\end{equation*}
where $\rho_k(f)$ is the lag $k$ auto-correlation of the Markov chain $\{f(\theta^{(1)}), \ldots, f(\theta^{(T)})\}$. Informally, the $\essmc$ of an MCMC output is interpretable as providing the equivalent number of independent simulations from the target distribution, where equivalent means with equivalent variance, as a function of the target acceptance probability $p_0$. 
Hence, the more efficient a sampler is in its
approximation to \textit{i.i.d.} sampling from the target, the larger the associated $\essmc$ should be.
In the following examples, we focus on the $\essmc$ for the estimation of the mean and the variance, \ie, $\essmc(\theta)$ and $\essmc(\theta^2)$, using the \texttt{monitor} function from the \texttt{Rstan} package.

\begin{figure}[t]%
\centering
\includegraphics[width=.49\textwidth]{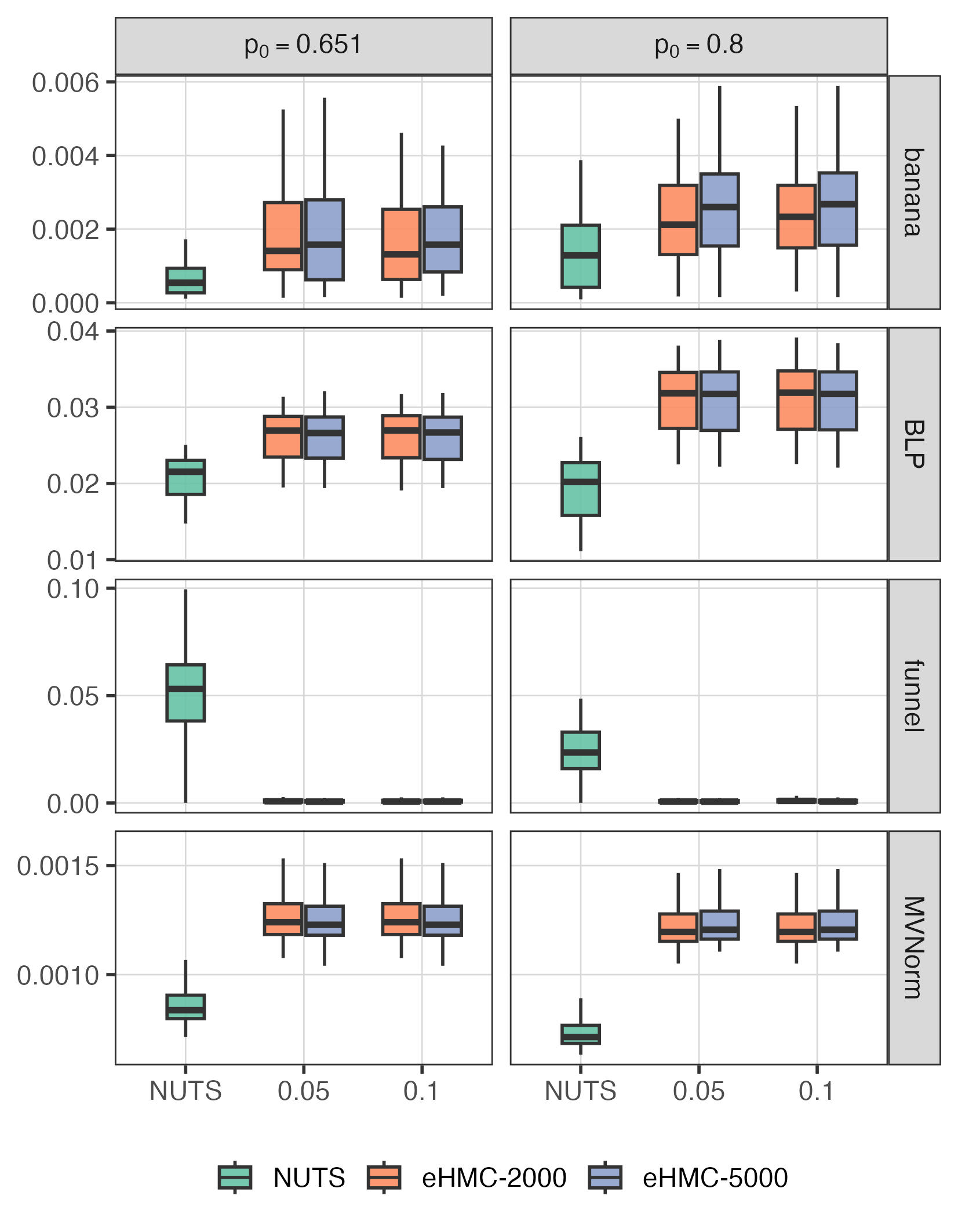}
\includegraphics[width=.49\textwidth]{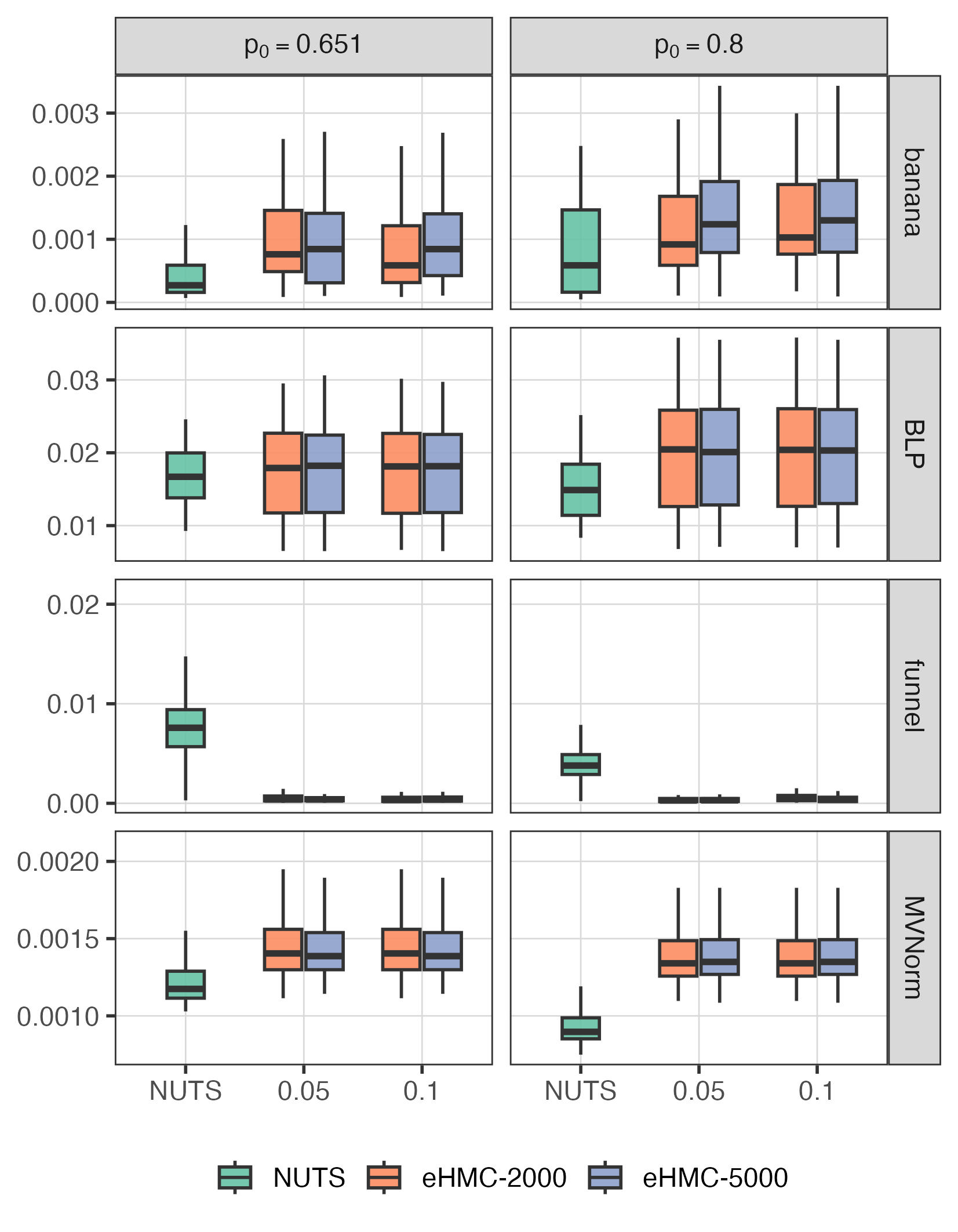}
\caption{Distribution of the Markov chain effective sample size per number of leapfrog steps for mean (left panel) and variance (right panel) estimation, aggregated across dimensions and 20 independent runs of 50 parallel chains of length $T = 2000$. 
For eHMC, the proposal is trained using $\lambda_0 = 0.8$, $N_0=50000$, from which $N \in \{2000, 5000\}$ resampled particles are used to construct the empirical measure on the integration parameters. 
Results are shown for $\Delta/N_0 \in \{5\%, 10\%\}$ in the $x$-axis.
}
\label{fig:ess-norm}
\end{figure}

\begin{figure}[t]%
\centering
\includegraphics[width=.49\textwidth]{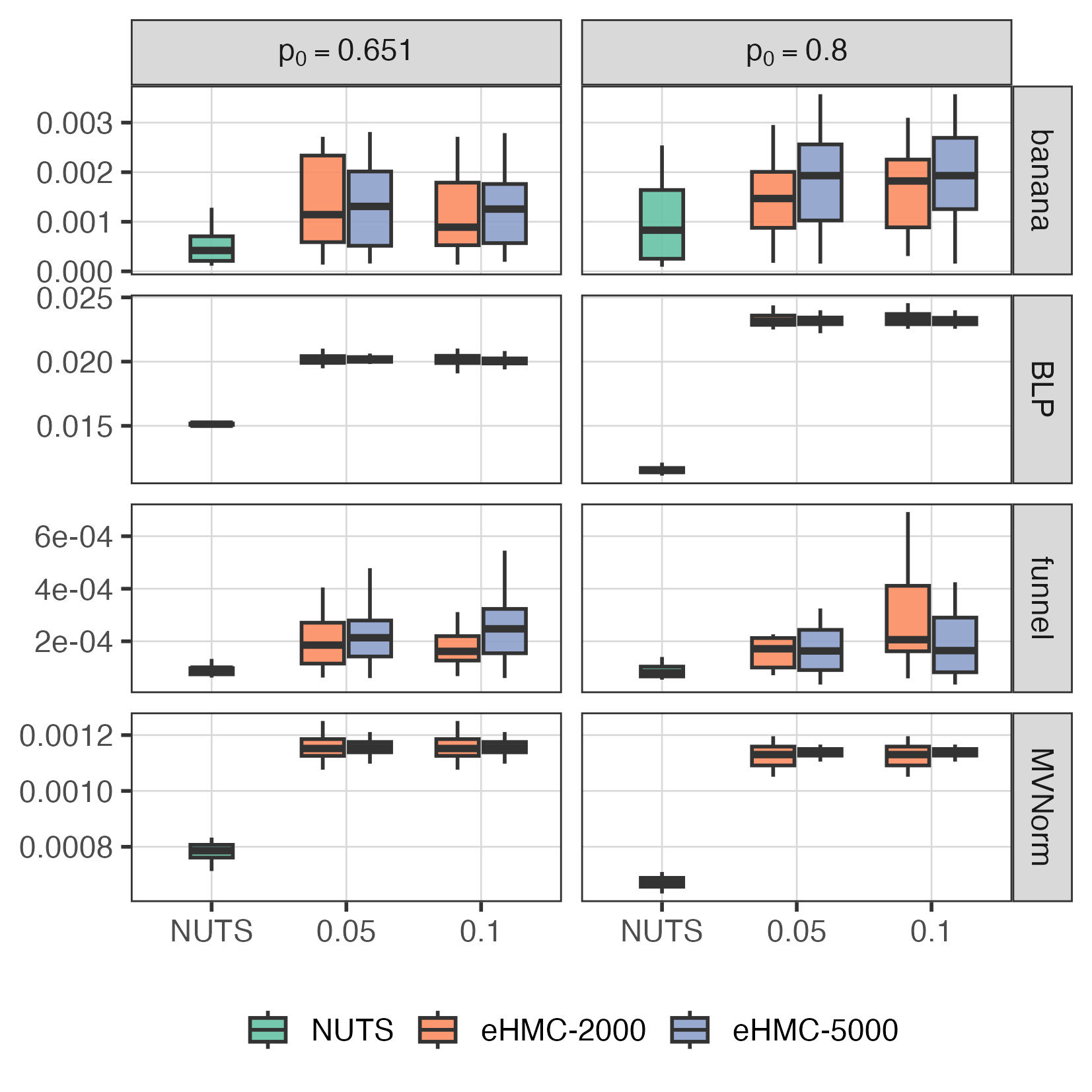}
\includegraphics[width=.49\textwidth]{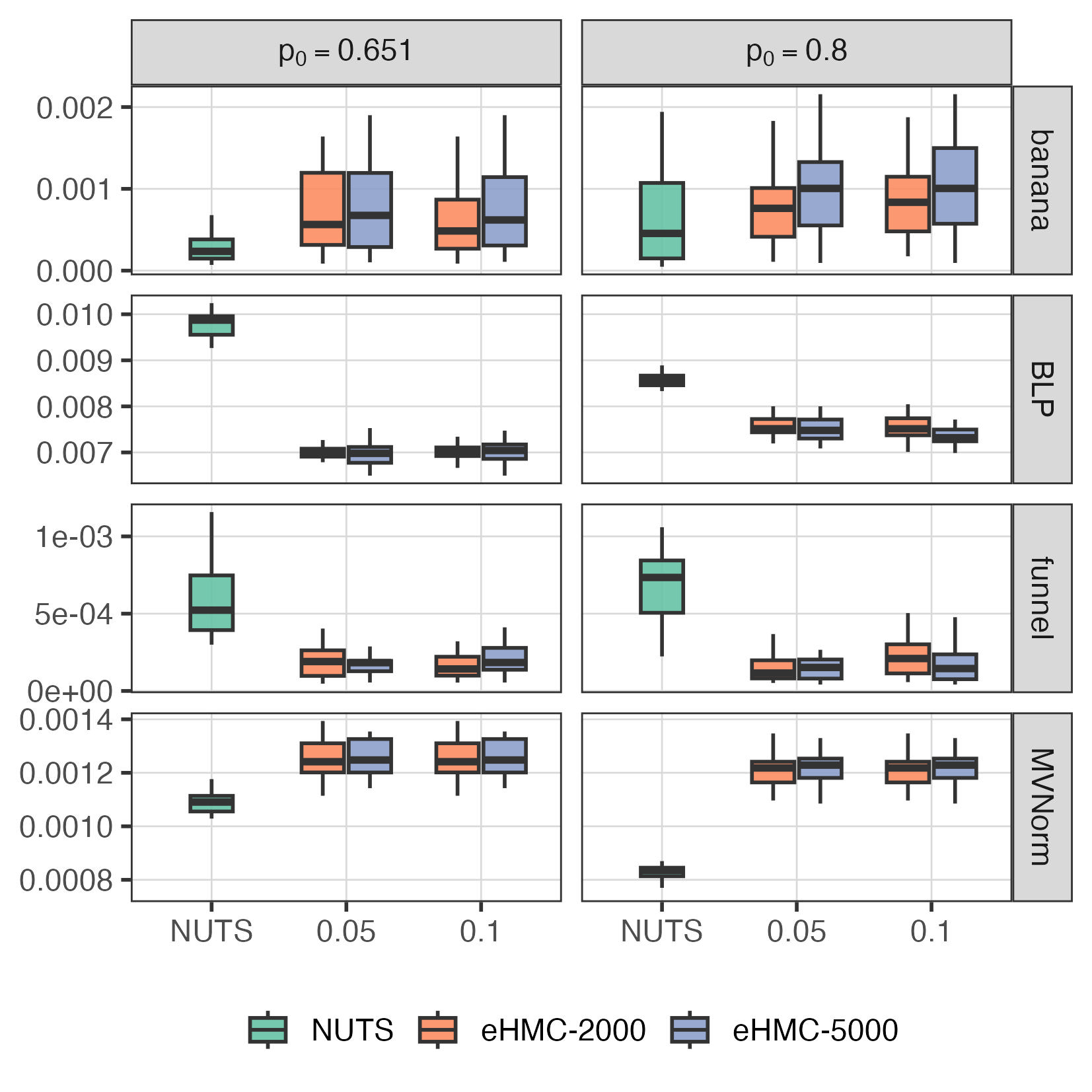}
\caption{Distribution of the minimum across dimensions of the Markov chain effective sample size per number of leapfrog steps for mean (left panel) and variance (right panel) estimation, aggregated over 20 independent runs of 50 parallel chains of length $T = 2000$. 
For eHMC, the proposal is trained using $\lambda_0 = 0.8$, $N_0=50000$, from which $N \in \{2000, 5000\}$ resampled particles are used to construct the empirical measure on the integration parameters. 
Results are shown for $\Delta/N_0 \in \{5\%, 10\%\}$ in the $x$-axis.
}
\label{fig:min-ess-norm}
\end{figure}

Figure \ref{fig:ess-norm} displays, for all configurations, the distribution of $\essmc(\theta)$ and $\essmc(\theta^2)$ per gradient evaluation across dimensions and runs. Figure \ref{fig:min-ess-norm} reports, over runs, the distribution of their minimum taken over all components of the parameter of the model.
This metric captures the least efficient direction of the sampler (see  Table \ref{tab:min-ess-mean} and Table \ref{tab:min-ess-var} in Appendix \ref{sec:add-result} for numerical summaries).

For the \textsf{banana}, \textsf{MVNorm}, and \textsf{BLP} targets, eHMC generally yields larger $\essmc(\theta)$ per gradient evaluation than NUTS across the configurations considered, with only modest sensitivity to $N$ and $\Delta/N_0$. The largest improvements are observed for the \textsf{banana} distribution, where eHMC achieves a median $\essmc(\theta)$ approximately 2 times larger than NUTS, while it is around 1.5 times larger for both \textsf{BLP} and \textsf{MVNorm} targets. 
On these model, between the configurations, eHMC generally provides a modest but consistent additional gain when $p_0$ increases. 
Furthermore, Figure~\ref{fig:min-ess-norm} shows that for these models eHMC exhibits improved sampling efficiency even along the least efficient direction of the sampler, although we observe a higher variability for the \textsf{banana} target.
The funnel distribution exhibits a more nuanced pattern. Although eHMC retains a larger minimum coordinate-wise $\essmc(\theta)$ for mean estimation than NUTS (see Table \ref{tab:min-ess-mean}), this metric does not capture movement between narrower and broader parts of the target (see results on ESJD). Overall, NUTS outperforms eHMC in terms of $\essmc(\theta)$ on this example. Increasing $p_0$ produces only modest and non-uniform changes across the funnel experiments.

Results for variance estimation are slightly more nuanced.
For the \textsf{banana}, \textsf{MVNorm}, and \textsf{BLP} targets, although eHMC achieves competitive or improved efficiency compared to NUTS in terms of the median $\essmc(\theta^2)$ when accounting for the number of leapfrog steps performed, performance in the least efficient direction of the sampler deteriorates for the \textsf{BLP} target, as illustrated in Figure~\ref{fig:min-ess-norm}. 
The contrast is stronger for the funnel example: NUTS yields a minimum variance ESS approximately three to four times larger than eHMC, depending on $p_0$ and the calibration configuration. Overall NUTS provides an $\essmc(\theta^2)$ that overseeds eHMC performance on this multiscale target.

\paragraph*{The expected squared jump distance (ESJD)} Furthermore, considering that the $\essmc$ criterion only reflects on the marginal efficiency of a sampler, we report in addition the standard expected squared jump distance. 
While depending on the measure of the jump, this quantity appears as a global mixing assessment that expresses another aspect of the efficiency of a sampler. We recall that, given generated samples $\{\theta^{(1)}, \ldots, \theta^{(T)}\}$, its estimated ESJD is defined as
\begin{equation*}
\text{EJSD} = \frac{1}{T-1}\sum_{t=1}^{T-1}\left\lVert \theta^{(t+1)} - \theta^{(t)}\right\rVert_2^2.
\end{equation*}

Table \ref{tab:esjd} summarizes the ESJD per gradient evaluation averaged over 20 independent runs (see Figure~\ref{fig:esjd-norm} in Appendix~\ref{sec:add-result} for the full distribution).
For the \textsf{banana}, \textsf{MVNorm}, and \textsf{BLP} targets, eHMC yields larger ESJD per gradient evaluation than NUTS. The funnel shows the opposite behavior: NUTS has an ESJD approximately 54 times larger for $p_0=0.651$ and 19 times larger for $p_0=0.8$ than the best-performing eHMC configuration. 
The funnel example shows a substantially larger ESJD for NUTS, approximately 54 times larger for $p_0=0.651$ and 19 times larger for $p_0=0.8$, than the best-performing eHMC configuration.
This indicates that eHMC makes much shorter effective moves through the state space. This behavior is consistent with the strong spatial heterogeneity of the funnel geometry, where the integration lengths required for efficient exploration vary considerably across the distribution support. Consequently, a single offline distribution of integration lengths is less able to accommodate these local changes than the online adaptation performed by NUTS.

\begin{table}[!b]
\caption{Expected squared jump distance averaged over 20 independent runs of 50 parallel chains of length $T = 2000$. For eHMC, the proposal is trained using $\lambda_0 = 0.8$, $N_0=50000$, from which $N$ resampled particles are used to construct the empirical measure on the integration parameters. Values are reported as mean $\pm 2$ standard errors, scaled by a factor $10^{6}$. \label{tab:esjd}}
\centering
\begin{tabular*}{\textwidth}{@{\extracolsep{\fill}}lccccc}
\toprule \\
 &  & \multicolumn{4}{c}{eHMC} \\
\cmidrule(lr){3-6}
Model & NUTS 
& \multicolumn{2}{c}{$\Delta/N_0 = 5\%$} 
& \multicolumn{2}{c}{$\Delta/N_0 =  10\%$} \\
\cmidrule(lr){3-4} \cmidrule(lr){5-6}
 &  & $N=2000$ & $N=5000$ & $N=2000$ & $N=5000$ \\
\midrule
\textsf{banana}   
& $44.2 \pm 2.7$ 
& $54.6 \pm 2.9$ 
& $54.4 \pm 2.6$ 
& $\mathbf{54.9 \pm 2.5}$
& $54.7 \pm 2.1$ 
\\
\textsf{MVNorm} 
& $3.88 \pm 0.062$
& $5.59 \pm 0.094$ 
& $\mathbf{5.60 \pm 0.12}$ 
& $5.59 \pm 0.094$ 
& $\mathbf{5.60 \pm 0.12}$ 
\\
\textsf{BLP}
& $0.742  \pm 0.0053$
& $1.15 \pm 0.026$ 
& $1.15 \pm 0.019$ 
& $\mathbf{1.16 \pm 0.016}$ 
& $1.15 \pm 0.018$
\\
\textsf{funnel}
& $\mathbf{1610  \pm 930}$
& $30.1 \pm 42$ 
& $27.8 \pm 47$ 
& $28.3 \pm 40$ 
& $27 \pm 50$
\\
\bottomrule
\end{tabular*}
\footnotetext{(a) Results for a target Metropolis--Hastings acceptance probability $p_0 = 0.651$.}

\vspace{0.5em}

\begin{tabular*}{\textwidth}{@{\extracolsep{\fill}}lccccc}
\toprule \\
 &  & \multicolumn{4}{c}{eHMC} \\
\cmidrule(lr){3-6}
Model & NUTS 
& \multicolumn{2}{c}{$\Delta/N_0 = 5\%$} 
& \multicolumn{2}{c}{$\Delta/N_0 =  10\%$} \\
\cmidrule(lr){3-4} \cmidrule(lr){5-6}
 &  & $N=2000$ & $N=5000$ & $N=2000$ & $N=5000$ \\
\midrule
\textsf{banana}   
& $37.7 \pm 2.9$ 
& $\mathbf{58.6 \pm 3.5}$ 
& $58.2 \pm 2.9$ 
& $58.3 \pm 2.6$ 
& $58.0 \pm 2.1$ 
\\
\textsf{MVNorm} 
& $3.2 \pm 0.064$
& $5.27 \pm 0.093$ 
& $\mathbf{5.27 \pm 0.080}$
& $5.27 \pm 0.093$ 
& $\mathbf{5.27 \pm 0.080}$ 
\\
\textsf{BLP}
& $0.648 \pm 0.012$
& $1.19 \pm 0.024$
& $\mathbf{1.19 \pm 0.015}$
& $1.19 \pm 0.017$
& $1.19 \pm 0.016$
\\
\textsf{funnel}
& $\mathbf{599  \pm 310}$
& $31.6 \pm 44$ 
& $28.1 \pm 49$ 
& $27.8 \pm 39$ 
& $25.1 \pm 46$
\\
\bottomrule
\end{tabular*}
\footnotetext{(b) Results for a target Metropolis--Hastings acceptance probability $p_0 = 0.8$.}
\end{table}

Taken together, the experiments expose settings in which our approach is useful. On the \textsf{banana}, \textsf{MVNorm}, and \textsf{BLP} examples, our empirical distribution of integration parameters yields higher sampling efficiency than NUTS for several of the considered metrics. The funnel provides a contrasting result. Because the appropriate integration scale varies strongly with position, the empirical distribution cannot reproduce the benefits of local trajectory adaptation, and NUTS performs substantially better in terms of both sampling efficiency and global movement of the Markov chain. In conclusion, and rather unsurprisingly, eHMC constitutes a relevant, automated, approach to randomized HMC that is well suited to targets whose relevant integration scales can be learned and reused offline. However, it is not intended to replace locally adaptive methods such as NUTS in settings where the appropriate integration time varies substantially across the state space.

\section{Extensions}

While the proposed approach focuses on learning a distribution over integration parameters to improve sampling efficiency, it can be naturally extended in several directions. In particular, we could introduce a partial refreshment of the momentum variable $v$, as commonly considered in persistent HMC methods \citep{neal2001}. 
Rather than fully resampling the momentum at each iteration, the idea is to retain it with probability $\eta \in [0,1]$, thereby inducing correlation between successive trajectories and potentially improving exploration efficiency. 
This partial refreshment also enables trajectory reuse and can be combined with Rao--Blackwellization strategies. 
This modification can be exploited in the empirical distribution of integration parameters. The resulting algorithm is referred to as prHMC.  

More formally, for any $(\varepsilon, L)$, let $P_{\varepsilon,L}$ denote the HMC transition kernel with step size $\varepsilon$ and $L$ leapfrog steps, and let $\widehat{\mu}_{\mathcal{T}}$ be a probability measure on $(0,\infty)\times\mathbb{N}$ such as \eqref{eqn:dist-time}. 
We then introduce a refreshment kernel that updates the auxiliary variables $(v,\varepsilon,L)$. 
At each iteration, the momentum is refreshed with probability $\eta \in [0,1]$. Two variants can be considered. 
In the first variant, the integration parameters $(\varepsilon, L)$ are kept fixed as long as the momentum is not refreshed. The corresponding kernel writes as
\begin{equation*}
R_\eta^{(1)}\left((v,\varepsilon,L), \dd v' \dd\varepsilon' \dd L'\right)
=
(1-\eta) \delta_{(v,\varepsilon,L)}(\dd v' \dd\varepsilon' \dd L')
+
\eta \mathcal{N}(0,M)(\dd v') \widehat{\mu}_{\mathcal{T}}(\dd\varepsilon', \dd L').
\end{equation*}

In the second variant, the step size $\varepsilon$ is kept fixed when the momentum is not refreshed, while the number of leapfrog steps $L$ is resampled at each iteration according to the marginal empirical distribution
\begin{equation*}
    \widehat{\mu}^\star_{\mathcal{L}} = \frac{1}{2}\sum_{i = 1}^N \omega_i \delta_{L(\vartheta_i, v_i, \varepsilon_i^+)} 
    + \frac{1}{2}\sum_{i = 1}^N \omega_i \delta_{L(\vartheta_i, -v_i, \varepsilon_i^-)},
\end{equation*}
while the step size $\varepsilon$ is resampled only when the momentum is refreshed.
The corresponding kernel then writes as
\begin{equation*}
R_\eta^{(2)}\left((v,\varepsilon,L), \dd v' \dd\varepsilon' \dd L'\right)
=
(1-\eta) \delta_v(\dd v') \delta_\varepsilon(\dd \varepsilon') \widehat{\mu}_{\mathcal{L}}^\star(\dd L')
+
\eta \mathcal{N}(0,M)(\dd v') \widehat{\mu}_{\mathcal{T}}(\dd\varepsilon', \dd L').
\end{equation*}

Keeping the step size fixed when the momentum is not refreshed enables the reuse of points previously computed along the discretized trajectory. This is achieved thanks to a flipping mechanism: the momentum is negated when a HMC proposal is rejected and left unchanged otherwise.
\begin{equation*}
Q\left((\theta,v), \dd \theta' \dd v' \right)
=\delta_\theta(\dd \theta')  \delta_{-v}(\dd v').
\end{equation*}
As a result, the sampler can explore a given level set both forward and backward, thereby reusing previously generated points when the direction of the momentum flips.

The prHMC kernel is then defined for any $\eta\in[0,1]$ as the mixture kernel obtained by composing the refreshment kernel and the flipping kernel with the HMC kernel, \ie,
\begin{equation*}
P_\eta^{(i)}((\theta,v,\varepsilon,L), \dd \theta' \dd v' \dd\varepsilon' \dd L')
=
\int
Q((\bar\theta,\bar v),\dd\theta'\dd v')
P_{\varepsilon',L'}\left((\theta,\tilde v), \dd \bar\theta \dd \bar v\right)
R_\eta^{(i)}\left((v,\varepsilon,L), \dd \tilde v \dd\varepsilon' \dd L'\right).
\end{equation*}
Since the Gaussian momentum distribution is invariant under the map $v \mapsto -v$, the flipping kernel $Q$ preserves the augmented target $\pi\otimes\mathcal{N}(0,M)$. Moreover, the refreshment kernel $R_\eta^{(i)}$ preserves $\mathcal{N}(0,M)\otimes\widehat{\mu}_{\mathcal T}$ by construction. 
In conclusion, since the HMC kernel leaves $\pi\otimes\mathcal{N}(0,M)$ invariant, the prHMC kernel leaves the augmented measure $\pi \otimes \mathcal{N}(0,M) \otimes \widehat{\mu}_{\mathcal T}$
invariant.

While this extension is theoretically appealing and naturally complements the proposed framework by leveraging intermediate proposals
along the leapfrog path to reduce the computational cost, the empirical results obtained on the set of examples considered in this paper remain inconclusive in terms of sampling efficiency (Figures \ref{fig:prHMCF-ess-norm} and \ref{fig:prHMCT-ess-norm} in Appendix \ref{sec:add-result}).
In particular, significant improvements are observed only on the \textsf{banana} distribution (Figures \ref{fig:prHMCF-ess-norm}).
In terms of ESJD, performances slightly deteriorate compared with eHMC, although it remains superior to NUTS (Figure \ref{fig:prHMC-esjd-norm} in Appendix \ref{sec:add-result}).
This suggests that the benefits of partial refreshment may depend sensitively on the geometry of the target distribution and the interaction with the learned integration parameters. Further investigation is therefore required to better understand the regimes in which this approach is most effective.

\section*{Code availability}
The code used in this article is available at \url{https://github.com/jstoehr/eHMC}.

\section*{Acknowledgements}

The early version of this paper was a component of Changye Wu's PhD thesis at Université Paris Dauphine, with which he is no longer affiliated. Christian P Robert is partially supported by the European Union under the ERC Synergy Grant 101071601 (OCEAN, 2023–2030). Views and opinions expressed are solely those of the authors and do not necessarily reflect those of the European Union or the European Research Council Executive Agency. Neither the European Union nor the granting authority can be held responsible for them. Christian P Robert is also partially supported by a PR[AI]RIE-PSAI Chair funded by the Agence Nationale de la Recherche (ANR-23-IACL-0008). Part of this paper was written while Christian P Robert was a visiting professor at Ca' Foscari University of Venice.
\bibliographystyle{abbrvnat} 
\bibliography{biblio}

\begin{appendices}

\section{Finitness of the U-turn condition}
\label{sec:finite-u-turn}

\paragraph*{Notation} In what follows, let denote $(\theta(t), v(t))$ the solution of the Hamilton's equations \eqref{eqn:hamilton-dyn} at time $t$.

\subsection{Preliminary result}

\begin{lemma}
\label{lem:finite-turning-time}
Assume that the target distribution $\pi$ admits a continuously differentiable density on $\mathbb{R}^d$, and that the potential $U : \theta \mapsto -\log \widetilde\pi(\theta)$. Then, given $(\theta(0), v(0)) \sim \pi \otimes \mathcal{N}(0,M)$, there exists $t^\star > 0$ such that the continuous-time Hamiltonian trajectory at time $t^\star$ satisfies
\begin{equation*}
(\theta(t^\star)-\theta(0))^\top M^{-1}v(t^\star) < 0,
\end{equation*}
almost surely.
\end{lemma}

\begin{proof}
In continuous-time Hamiltonian dynamics, the conservation of the Hamiltonian and coercivity of $U$ yield that the level sets $\{(\theta,v) : H(\theta,v) \le c\}$ are compact. Otherwise, given $v\in\mathbb{R}^d$, we could find a sequence $(\theta_n)_{n\ge 1}$ such that $\lim_{n\to\infty} \theta_n = \infty$ and for all $n\ge 1$, $U(\theta_n) \le c$, which contradicts the coercivity assumption. 
Consequently, for all $t\ge 0$, the trajectory $(\theta(t), v(t))$ remains in a compact set $\mathcal{C}$.

Let define for all $t\ge 0$, $g(t) = \lVert \theta(t) - \theta(0)\rVert_2^2$. Then, using Hamiltonian's equations,
\begin{equation*}
g'(t) = 2(\theta(t)-\theta(0))^\top M^{-1}v(t).
\end{equation*}
Hence, to prove the result, we establish that $g$ is a decreasing function.

Assume that, for all $t \ge 0$, $g'(t) \ge 0$. Then $g$ is non-decreasing, and, since $(\theta(t), v(t))$ remain in a compact set, $g(t)$ is bounded. It follows that $\lim_{t\to\infty}g(t)$ is finite.
Since the Hamiltonian flow preserves the Lebesgue measure and the Lebesgue measure is finite on the compact set $\mathcal{C}$, the Poincaré recurrence theorem implies that, for almost every initial condition $(\theta(0), v(0))$, the trajectory returns arbitrarily close to its starting point infinitely often, \ie, there exists a sequence $(t_n)_{n\ge 0}$ such that
\begin{equation*}
\lim_{n \to \infty} t_n = \infty, \qquad
\lim_{n \to \infty} (\theta(t_n), v(t_n)) = (\theta(0), v(0)).
\end{equation*}
It follows, in particular,  that $\lim_{n \to \infty} g(t_n) = 0$. 

Since $g$ is non-decreasing and $g(0)=0$, this implies that for all $t\ge 0$, $g(t) = 0$. Thus, for all $t\ge 0$, we have $\theta(t) = \theta(0)$. By the first Hamiltonian equation, we get that for all $t \ge 0$, $v(t) = 0$. 
Therefore, the trajectories for which $(\theta(t),v(t))$ remains constant belong to the set of trajectories such that $v(0) = 0$, a set with zero measure under $\pi \otimes \mathcal{N}(0,M)$.

In conclusion, for almost every initial condition, there exists $t^\star > 0$ such that
\begin{equation*}
\frac{1}{2}g'(t^\star) = (\theta(t^\star)-\theta(0))^\top M^{-1}v(t^\star) < 0.
\end{equation*}
\end{proof}

\subsection{Proof of Lemma \ref{lem:finite-turning-time-discrete}}

Let define for all $(\theta, v)$, and all $t \ge 0$,
\begin{equation*}
g(t) = (\theta(t)-\theta)^\top M^{-1} v(t^\star).
\end{equation*}
From Lemma \ref{lem:finite-turning-time}, for $\pi \otimes \mathcal{N}(0,M)$-almost every initial condition $(\theta,v)$, there exists a finite time $t^\star > 0$ such that $g(t^\star) < 0$.
Since $g$ is continuous, for all $\alpha < 0$, there exists $\eta > 0$ such that
\begin{equation}
\label{eqn:g-bound}
\forall t\in(t^\star-\eta, t^\star+\eta), \quad g(t) \le \alpha.
\end{equation}

Let $\alpha < 0$. Given $\varepsilon > 0$, the leapfrog integrator approximates the true solution with error $\mathcal{O}(\varepsilon^2)$ on any fixed finite time interval \citep{leimkuhler2005}, \ie, there exists a constant $C_1>0$ such that for all $\ell$ with $\varepsilon\ell \le t^\star + \eta$
\begin{equation*}
\left\lVert (\theta_\ell, v_\ell) - (\theta(\varepsilon \ell), v(\varepsilon\ell))\right\rVert_2 \le C_1 \varepsilon^2, \qquad (\theta_\ell, v_\ell) = F_\varepsilon^\ell(\theta, v).
\end{equation*}
Moreover, in the proof of Lemma \ref{lem:finite-turning-time} we have seen that the level sets of the Hamiltonian $\{(\theta,v) : H(\theta,v) \le c\}$ are compact.
From the approximation error of the leapfrog integrator, there exists $C_2>0$ such that for all $\ell$ with $\varepsilon\ell \le t^\star + \eta$,
\begin{equation*}
\left| H(\theta_\ell, v_\ell) - H(\theta, v) \right| \le C_2 \varepsilon^2.
\end{equation*}
It follows that
\begin{equation*}
(\theta_\ell, v_\ell) \in \left\{ (\theta',v') : H(\theta',v') \le H(\theta,v) + C_2\varepsilon^2 \right\}.
\end{equation*}
which is compact.
The function $\psi : (\widetilde\theta, v) \mapsto (\widetilde\theta-\theta)^\top M^{-1}v$ is infinitely differentiable on $\mathbb{R}^{2d}$ and therefore Lipschitz on any compact set. We thus have that there exists a constant $C_3$ such that for all $\ell$ with $\varepsilon\ell \le t^\star + \eta$
\begin{equation}
\label{eqn:lipschitz-bound}
\left\lvert \psi(\theta_\ell, v_\ell) - g(\varepsilon\ell) \right\rvert \le C_3\left\lVert (\theta_\ell, v_\ell) - (\theta(\varepsilon \ell), v(\varepsilon\ell))\right\rVert_2 \le C_1C_3 \varepsilon^2.
\end{equation}

Let set 
\begin{equation*}
\varepsilon_0 = \sqrt{\frac{-\alpha}{C_1C_3}}.
\end{equation*}
For all $\varepsilon \leq \varepsilon_0$, it follows from \eqref{eqn:g-bound} and \eqref{eqn:lipschitz-bound} that for all $\ell$ such that $\lvert \varepsilon\ell - t^\star \rvert \le \eta$
\begin{equation}
\label{eqn:cond-return}
\psi(\theta_\ell, v_\ell) = (\theta_\ell-\theta)^\top M^{-1}v_\ell 
\le C_1C_3 \varepsilon^2 + g(\varepsilon \ell)
\le 2\alpha < 0.
\end{equation}
To conclude, there exists $\varepsilon^\star \leq \varepsilon_0$ such that $\{\ell \in \mathbb{N} : \lvert \varepsilon^\star\ell - t^\star \rvert \le \eta\}$ is non empty. Thus, for all $\varepsilon \leq \varepsilon^\star$, there exists $\ell \in \mathbb{N}$ that satisfies \eqref{eqn:cond-return}, which implies that $L(\theta,v,\varepsilon) < \infty$.

\section{Formula for the mass matrix update}
\label{sec:formula}

Using the same notations as in Section \ref{sec:calibration}, let denote for $1\le i \le N$
\begin{align*}
L_i^+ & = L(\vartheta_i, v_i, \varepsilon_i^+),
&\rho_{i,\ell}^+ &= \rho(\vartheta_i, v_i, \vartheta_{i,\ell}^+, v_{i, \ell}^+),
\\
L_i^- & = L(\vartheta_i, -v_i, \varepsilon_i^-),
&\rho_{i,\ell}^- &= \rho(\vartheta_i, -v_i, \vartheta_{i,\ell}^-, v_{i, \ell}^-).
\end{align*}
The normalizing constant of the measure  $\nu$ as defined in \eqref{eqn:emp-measure-mass-matrix} is
\begin{equation*}
Z = \sum_{i=1}^N \omega_i \left(L_i^+ + L_i^-\right).
\end{equation*}

\paragraph*{Mean of the measure $\nu$}
\begin{equation}
\label{eqn:mean-nu}
m =
\frac{1}{Z}
\sum_{i=1}^N \omega_i \left[
\sum_{\ell=1}^{L_i^+}
\left\{
(1-\rho_{i,\ell}^+)\theta_{i,\ell-1}^+
+
\rho_{i,\ell}^+\theta_{i,\ell}^+
\right\}
+
\sum_{\ell=1}^{L_i^-}
\left\{
(1-\rho_{i,\ell}^-)\theta_{i,\ell-1}^-
+
\rho_{i,\ell}^-\theta_{i,\ell}^-
\right\}
\right],
\end{equation}

\paragraph*{Variance of the measure $\nu$} Let denote
\begin{align}
V_{i}^+ = \sum_{\ell=1}^{L_i^+} (1-\rho_{i,\ell}^+)(\theta_{i,\ell-1}^+ - m)(\theta_{i,\ell-1}^+ - m)^\top,
\qquad 
S_{i}^+ = \sum_{\ell=1}^{L_i^+} \rho_{i,\ell}^+(\theta_{i,\ell}^+ - m)(\theta_{i,\ell}^+ - m)^\top, \label{eqn:varp-nu}
\\
V_{i}^- = \sum_{\ell=1}^{L_i^-} (1-\rho_{i,\ell}^-)(\theta_{i,\ell-1}^- - m)(\theta_{i,\ell-1}^- - m)^\top,
\qquad 
S_{i}^- = \sum_{\ell=1}^{L_i^-} \rho_{i,\ell}^-(\theta_{i,\ell}^- - m)(\theta_{i,\ell}^+ - m)^\top. \label{eqn:varn-nu}
\end{align}
Then the variance is given by
\begin{equation}
\label{eqn:var-nu}
\mathrm{Var}_\nu(\theta)
=
\frac{1}{Z}
\sum_{i=1}^N \omega_i \left(V_i^+ + S_i^+ + V_i^- + S_i^- \right).
\end{equation}

\section{Invariance of the randomized kernel}

\begin{lemma}
\label{lemma:invariance}
Let $\widehat{\mu}_{\mathcal{T}}$ be a probability measure on $(0,\infty)\times\mathbb{N}$ constructed during the calibration stage and kept fixed during the sampling phase. For any $(\varepsilon, L)$, let $P_{\varepsilon,L}$ denote the Hamiltonian Monte Carlo transition kernel with step size $\varepsilon$ and $L$ leapfrog steps. Define the Markov transition kernel
\begin{equation*}
P(x, \dd x') = \int P_{\varepsilon,L}(x, \dd x')  \widehat{\mu}_{\mathcal{T}}(\dd\varepsilon, \dd L).
\end{equation*}
Then $P$ leaves $\pi$ invariant.
\end{lemma}

\begin{proof}
For any fixed $(\varepsilon, L)$, the Hamiltonian Monte Carlo kernel $P_{\varepsilon,L}$ preserves the target distribution $\pi$, that is
\begin{equation*}
\int \pi(\dd x) P_{\varepsilon,L}(x, \dd x') = \pi(\dd x').
\end{equation*}
Then, by Fubini's theorem,
\begin{align*}
\int \pi(\dd x) P(x, \dd x') 
&= \int \pi(\dd x) \int P_{\varepsilon,L}(x, \dd x')  \widehat{\mu}_{\mathcal{T}}(\dd\varepsilon, \dd L) \\
&= \int \left( \int \pi(\dd x) P_{\varepsilon,L}(x, \dd x') \right) \widehat{\mu}_{\mathcal{T}}(\dd\varepsilon, \dd L) \\
&= \int \pi(\dd x')  \widehat{\mu}_{\mathcal{T}}(\dd\varepsilon, \dd L) \\
&= \pi(\dd x').
\end{align*}
Therefore $\pi$ is invariant for $P$.
\end{proof}

\section{Monotonicity of the effective sample size}
\label{sec:monotonicity}

\begin{lemma}
\label{lem:ess-monotone}
Let $\mu\in\mathbb{M}_\pi$. Assume that for all $\beta \in [0,1]$, $\pi/\mu \in L^{2\beta}(\mu)$. Then, the function $\beta \mapsto \ess\left( \widetilde\pi^{\beta} \mu^{1-\beta}, \mu\right)$ is continuous and non-increasing on $[0,1]$.
\end{lemma}

\begin{proof}
Assume $\vartheta \sim \mu$. Let define the random variable $X = \log(\pi(\vartheta)/\mu(\vartheta))$. Then we can write
\begin{equation*}
\ess\left( \widetilde\pi^{\beta} \mu^{1-\beta}, \mu\right)
=
\frac{\mathbb{E}_{\mu}[\exp(\beta X)]^2}
{\mathbb{E}_{\mu}[\exp(2\beta X)]}.
\end{equation*}

\paragraph*{Continuity} The mapping $\beta \mapsto \exp(\beta X)$ is continuous on $[0,2]$. Moreover, for all $\beta \in [0,2]$, $\exp(\beta X) \le 1 + \exp(2X)$.
Since, by assumption, $\mathbb{E}_\mu[\exp(2X)] < \infty$, we have that $1 + \exp(2X)$ is integrable with respect to $\mu$. Therefore, by the dominated convergence theorem, $\beta \mapsto \mathbb{E}_\mu[\exp(\beta X)]$ is continuous on $[0,2]$. It follows that $\beta \mapsto \ess\left( \widetilde\pi^{\beta} \mu^{1-\beta}, \mu\right)$ is continuous on $[0,1]$ as ratio of positive and continuous functions.

\paragraph*{Monotonicity} Define for all $\beta\in[0,1]$
\begin{equation*}
g(\beta) = \log \ess\left( \widetilde\pi^{\beta} \mu^{1-\beta}, \mu\right)
= 2 \kappa(b) - \kappa(2\beta),\qquad \kappa : \beta \mapsto \log \mathbb{E}_{\mu}[\exp(\beta X)].
\end{equation*}
The function $\kappa$ is the cumulant generating function associated with $X$. It is twice differentiable on the interior of the set
$\mathcal{D} = \{\beta\in\mathbb{R} : \mathbb{E}_{\mu}[\exp(\beta X)] < \infty \}$.
Since, for all $\beta \in [0,1]$, $\pi/\mu \in L^{2\beta}(\mu)$, we have that $[0, 2] \subseteq \mathcal{D}$. Thus $\kappa$ is twice differentiable on $(0, 2)$. If, for all $\beta\in(0,2)$, we define the exponentially tilted measure
\begin{equation*}
\mu_\beta(\dd \vartheta)=\frac{\exp(\beta X)}{\mathbb{E}_\mu[\exp(\beta X)}\mu(\dd\vartheta),
\end{equation*}
we have
\begin{equation*}
\kappa'(\beta) = \mathbb{E}_{\mu_\beta}[X], \qquad \kappa''(\beta) = \mathbb{V}\mathrm{ar}_{\mu_\beta}[X] \geq 0.
\end{equation*}
It follows that $\kappa'$ is non-decreasing on $(0,2)$ and for all $\beta \in (0, 1)$,
\begin{equation*}
g'(\beta) = 2 \kappa'(b) - 2\kappa'(2\beta) \leq 0.
\end{equation*}
Therefore, $g$ is non-increasing on $(0, 1)$. Since $g$ is continuous on $[0,1]$, we get that $g$, and consequently $\beta \mapsto \ess\left( \widetilde\pi^{\beta} \mu^{1-\beta}, \mu\right)$, is non-increasing on $[0, 1]$.
\end{proof}

\section{Additional numerical results}
\label{sec:add-result}

\begin{table}[!h]
\caption{Minimum across dimensions of the Markov chain effective sample size per leapfrog step, averaged over 20 independent runs of 50 parallel chains of length $T = 2000$, for the estimation of the mean. For eHMC, the proposal is trained using $\lambda_0 = 0.8$, $N_0=50000$, from which $N$ resampled particles are used to construct the empirical measure on the integration parameters. Values are reported as mean $\pm 2$ standard errors, scaled by a factor $10^{4}$. \label{tab:min-ess-mean}}
\centering

\begin{tabular*}{\textwidth}{@{\extracolsep{\fill}}lccccc}
\toprule \\
 &  & \multicolumn{4}{c}{eHMC} \\
\cmidrule(lr){3-6}
Model & NUTS 
& \multicolumn{2}{c}{$\Delta/N_0 = 5\%$} 
& \multicolumn{2}{c}{$\Delta/N_0 =  10\%$} \\
\cmidrule(lr){3-4} \cmidrule(lr){5-6}
 &  & $N=2000$ & $N=5000$ & $N=2000$ & $N=5000$ \\
\midrule
\textsf{banana}   
& $5.26 \pm 8.7$ 
& $\mathbf{13.6 \pm 18}$ 
& $13.1 \pm 18.4$ 
& $11.7 \pm 17$ 
& $12.8 \pm 16$ 
\\
\textsf{MVNorm} 
& $7.81 \pm 0.67$
& $\mathbf{11.6 \pm 0.91}$ 
& $11.5 \pm 0.87$ 
& $\mathbf{11.6 \pm 0.91}$ 
& $11.5 \pm 0.87$ 
\\
\textsf{BLP}
& $151  \pm 3.3$
& $\mathbf{202 \pm 8.0}$ 
& $202 \pm 8.1$ 
& $202 \pm 10$ 
& $201 \pm 7.9$
\\
\textsf{funnel}
& $0.902  \pm 0.38$
& $2.15 \pm 2.8$ 
& $2.35 \pm 2.7$ 
& $1.86 \pm 1.9$ 
& $\mathbf{2.63 \pm 2.9}$
\\
\bottomrule
\end{tabular*}
\footnotetext{(a) Results for a target Metropolis--Hastings acceptance probability $p_0 = 0.651$.}

\vspace{0.5em}

\begin{tabular*}{\textwidth}{@{\extracolsep{\fill}}lccccc}
\toprule \\
 &  & \multicolumn{4}{c}{eHMC} \\
\cmidrule(lr){3-6}
Model & NUTS 
& \multicolumn{2}{c}{$\Delta/N_0 = 5\%$} 
& \multicolumn{2}{c}{$\Delta/N_0 =  10\%$} \\
\cmidrule(lr){3-4} \cmidrule(lr){5-6}
 &  & $N=2000$ & $N=5000$ & $N=2000$ & $N=5000$ \\
\midrule
\textsf{banana}   
& $10.3 \pm 16$ 
& $15.2 \pm 16$ 
& $\mathbf{18.5 \pm 19}$ 
& $17.3 \pm 18$ 
& $19 \pm 19$ 
\\
\textsf{MVNorm} 
& $6.73 \pm 0.42$
& $11.3 \pm 0.81$ 
& $\mathbf{11.4 \pm 0.42}$ 
& $11.3 \pm 0.81$ 
& $\mathbf{11.4 \pm 0.42}$ 
\\
\textsf{BLP}
& $116  \pm 5.4$
& $232 \pm 10$ 
& $232 \pm 12$ 
& $\mathbf{234 \pm 11}$ 
& $232 \pm 9.3$
\\
\textsf{funnel}
& $0.887  \pm 0.63$
& $1.83 \pm 2.2$ 
& $1.89 \pm 2.8$ 
& $\mathbf{2.85 \pm 3.6}$ 
& $2.04 \pm 3.1$
\\
\bottomrule
\end{tabular*}
\footnotetext{(b) Results for a target Metropolis--Hastings acceptance probability $p_0 = 0.8$.}
\end{table}
 
\begin{table}[!h]
\caption{Minimum across dimensions of the Markov chain effective sample size per leapfrog step, averaged over 20 independent runs of 50 parallel chains of length $T = 2000$, for the estimation of the variance. For eHMC, the proposal is trained using $\lambda_0 = 0.8$, $N_0=50000$, from which $N$ resampled particles are used to construct the empirical measure on the integration parameters. Values are reported as mean $\pm 2$ standard errors, scaled by a factor $10^{4}$. \label{tab:min-ess-var}}
\centering
\begin{tabular*}{\textwidth}{@{\extracolsep{\fill}}lccccc}
\toprule \\
 &  & \multicolumn{4}{c}{eHMC} \\
\cmidrule(lr){3-6}
Model & NUTS 
& \multicolumn{2}{c}{$\Delta/N_0 = 5\%$} 
& \multicolumn{2}{c}{$\Delta/N_0 =  10\%$} \\
\cmidrule(lr){3-4} \cmidrule(lr){5-6}
 &  & $N=2000$ & $N=5000$ & $N=2000$ & $N=5000$ \\
\midrule
\textsf{banana}   
& $3.39 \pm 6.3$ 
& $7.29 \pm 10$ 
& $7.46 \pm 11$ 
& $6.36 \pm 9.8$
& $\mathbf{7.36 \pm 10}$ 
\\
\textsf{MVNorm} 
& $10.9 \pm 0.81$
& $12.5 \pm 1.6$ 
& $12.5 \pm 1.4$ 
& $12.5 \pm 1.3$ 
& $\mathbf{12.5 \pm 1.4}$ 
\\
\textsf{BLP}
& $\mathbf{97.9  \pm 5.5}$
& $70.3 \pm 4.8$ 
& $69.8 \pm 5.1$ 
& $70.5 \pm 5.1$ 
& $70.0 \pm 4.8$
\\
\textsf{funnel}
& $\mathbf{6.09  \pm 5.4}$
& $2.14 \pm 3.0$ 
& $1.9 \pm 1.9$ 
& $1.83 \pm 2.4$ 
& $2.29 \pm 3.0$
\\
\bottomrule
\end{tabular*}
\footnotetext{(a) Results for a target Metropolis--Hastings acceptance probability $p_0 = 0.651$.}

\vspace{0.5em}

\begin{tabular*}{\textwidth}{@{\extracolsep{\fill}}lccccc}
\toprule \\
 &  & \multicolumn{4}{c}{eHMC} \\
\cmidrule(lr){3-6}
Model & NUTS 
& \multicolumn{2}{c}{$\Delta/N_0 = 5\%$} 
& \multicolumn{2}{c}{$\Delta/N_0 =  10\%$} \\
\cmidrule(lr){3-4} \cmidrule(lr){5-6}
 &  & $N=2000$ & $N=5000$ & $N=2000$ & $N=5000$ \\
\midrule
\textsf{banana}   
& $6.61 \pm 13$ 
& $7.79 \pm 9.4$ 
& $\mathbf{10.1 \pm 11}$ 
& $9.28 \pm 11$ 
& $10.3 \pm 11$ 
\\
\textsf{MVNorm} 
& $8.28 \pm 0.69$
& $12.1 \pm 1.2$ 
& $\mathbf{12.2 \pm 1.2}$ 
& $12.1 \pm 1.2$ 
& $\mathbf{12.2 \pm 1.2}$ 
\\
\textsf{BLP}
& $\mathbf{85.7  \pm 3.2}$
& $75.4 \pm 5.7$ 
& $75.0 \pm 5.3$ 
& $75.7 \pm 6.0$ 
& $73.7 \pm 5.0$
\\
\textsf{funnel}
& $\mathbf{6.77  \pm 4.7}$
& $1.53 \pm 2.0$ 
& $1.57 \pm 2.0$ 
& $2.27 \pm 2.6$ 
& $1.92 \pm 3.0$
\\
\bottomrule
\end{tabular*}
\footnotetext{(b) Results for a target Metropolis--Hastings acceptance probability $p_0 = 0.8$.}
\end{table}

\begin{figure}[!h]%
\centering
\includegraphics[width=\textwidth]{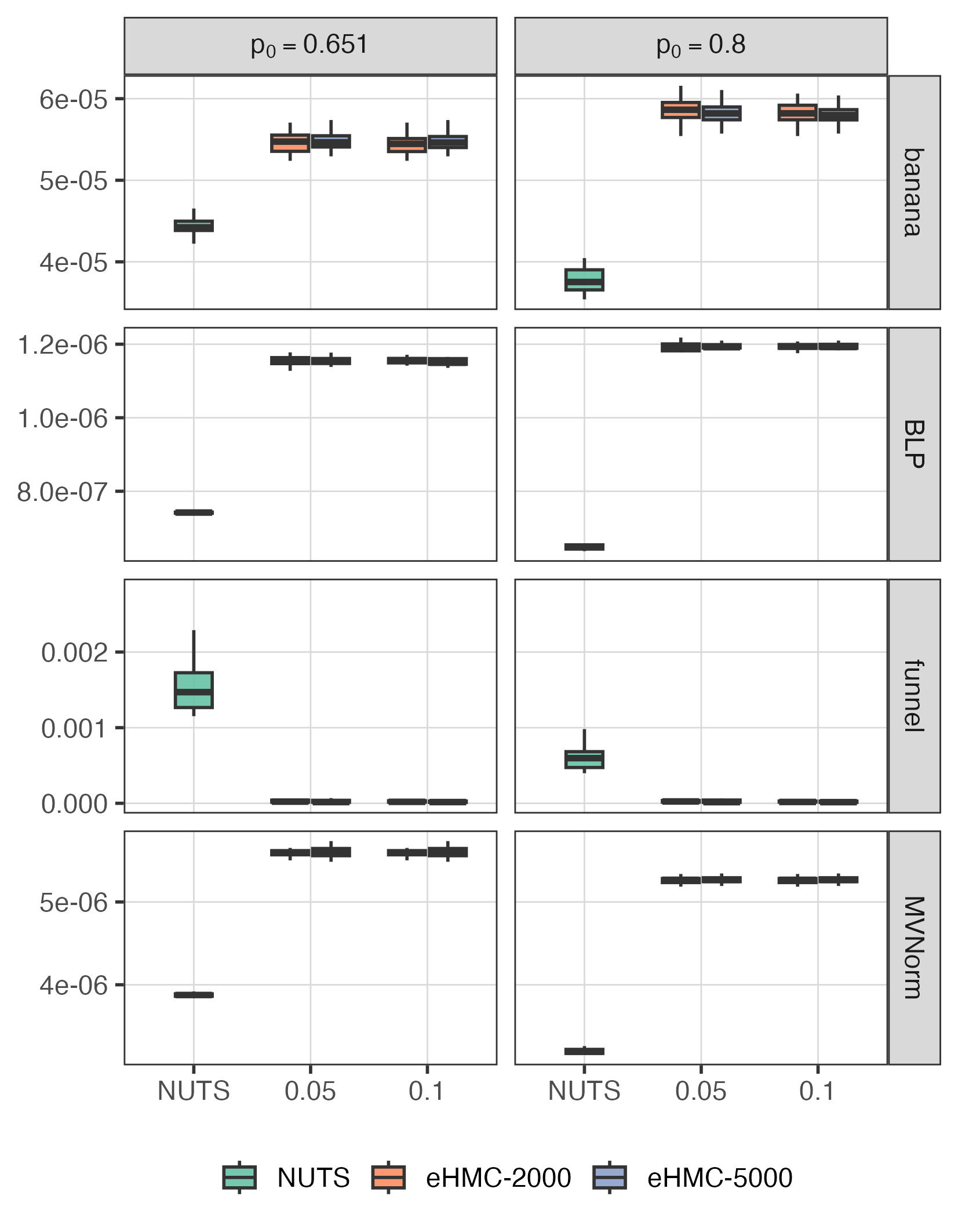}
\caption{Distribution of the expected squared jump distance aggregated over 20 independent runs of 50 parallel chains of length $T = 2000$. 
For eHMC, the proposal is trained using $\lambda_0 = 0.8$, $N_0=50000$, from which $N \in \{2000, 5000\}$ resampled particles are used to construct the empirical measure on the integration parameters. 
Results are shown for $\Delta/N_0 \in \{5\%, 10\%\}$ in the $x$-axis.}
\label{fig:esjd-norm}
\end{figure}

\begin{figure}[!h]%
\centering
\includegraphics[width=.49\textwidth]{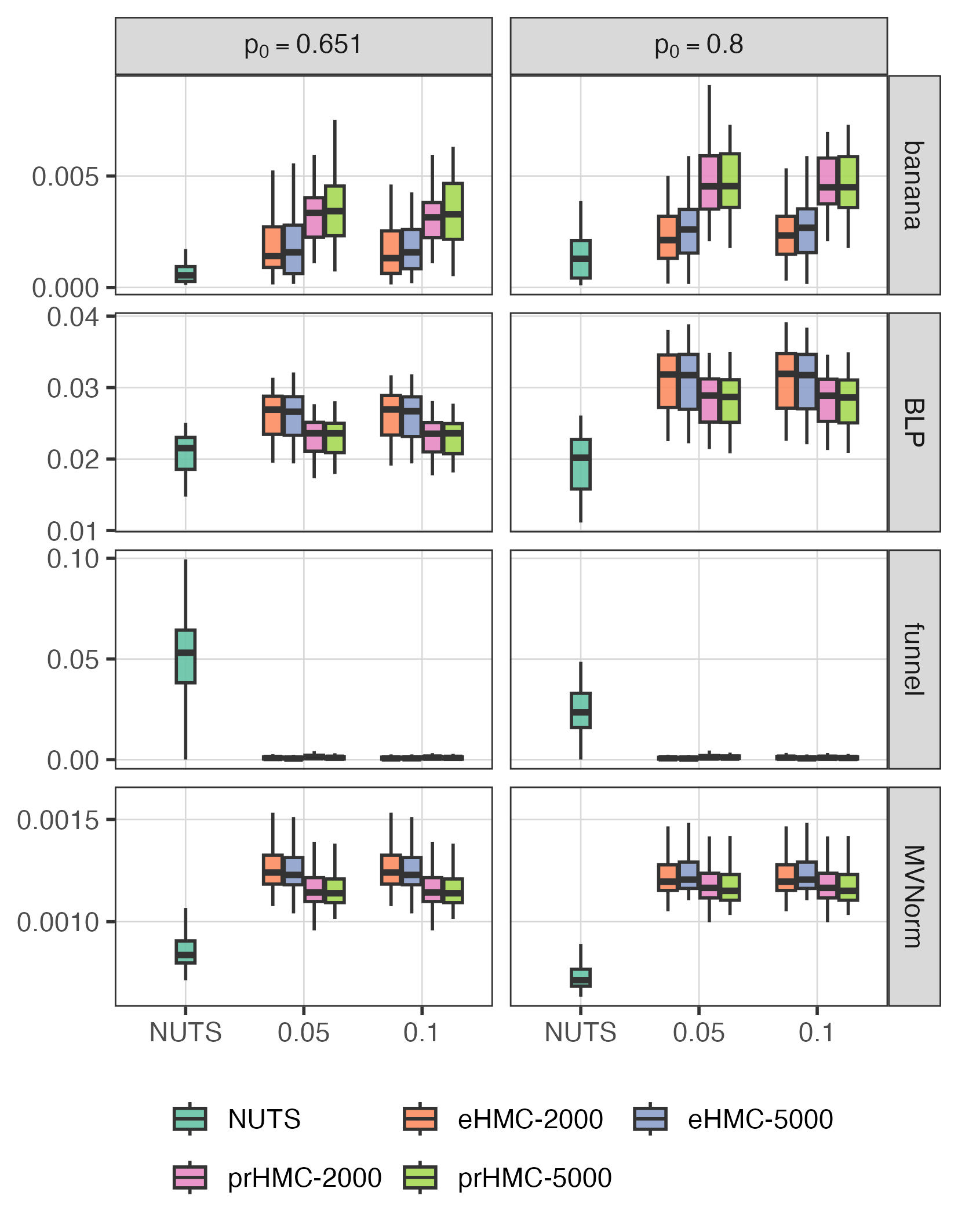}
\includegraphics[width=.49\textwidth]{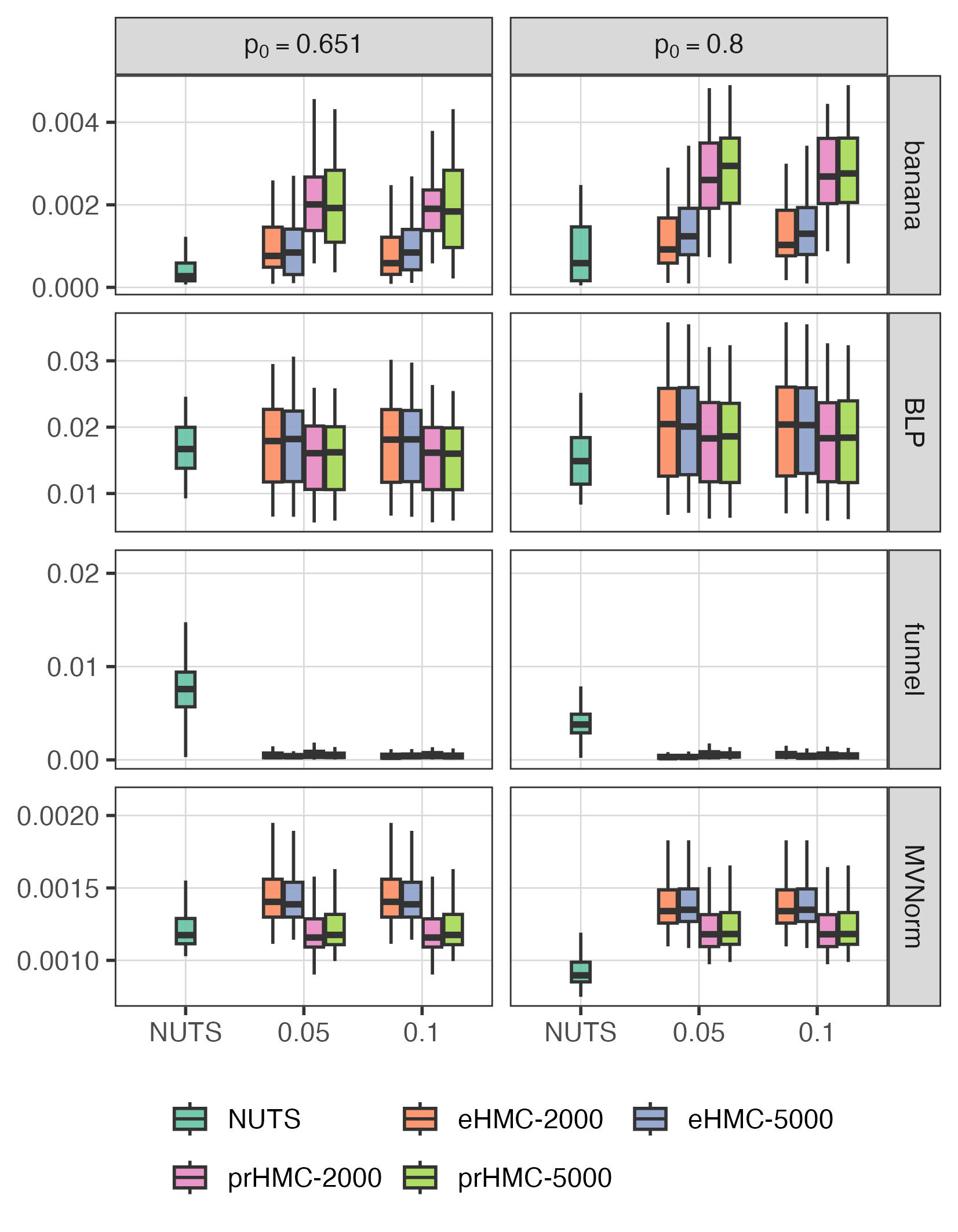}
\caption{Distribution of the Markov chain effective sample size per number of leapfrog steps for mean (left panel) and variance (right panel) estimation, aggregated across dimensions and 20 independent runs of 50 parallel chains of length $T = 2000$. 
prHMC denotes the variant $P_\eta^{(1)}$, where $(v, \varepsilon, L)$ are partially refreshed with probability $\eta = 0.75$. The same proposal is used for both eHMC and prHMC, trained with $\lambda_0 = 0.8$, $N_0 = 50000$ samples, from which $N \in \{2000, 5000\}$ particles are resampled to construct the empirical distribution on integration parameters. Results are shown for $\Delta/N_0 \in \{5\%, 10\%\}$ in the $x$-axis.
}
\label{fig:prHMCF-ess-norm}
\end{figure}

prHMC denotes the variant $P_\eta^{(1)}$, where $(v, \varepsilon, L)$ are partially refreshed with probability $\eta = 0.75$. The same proposal is used for both eHMC and prHMC, trained with $N_0 = 50000$ samples, from which $N \in \{2000, 5000\}$ particles are resampled to construct the empirical distribution on integration parameters. Results are shown for $\Delta/N_0 \in \{5\%, 10\%\}$ in the $x$-axis.

\begin{figure}[!h]%
\centering
\includegraphics[width=.49\textwidth]{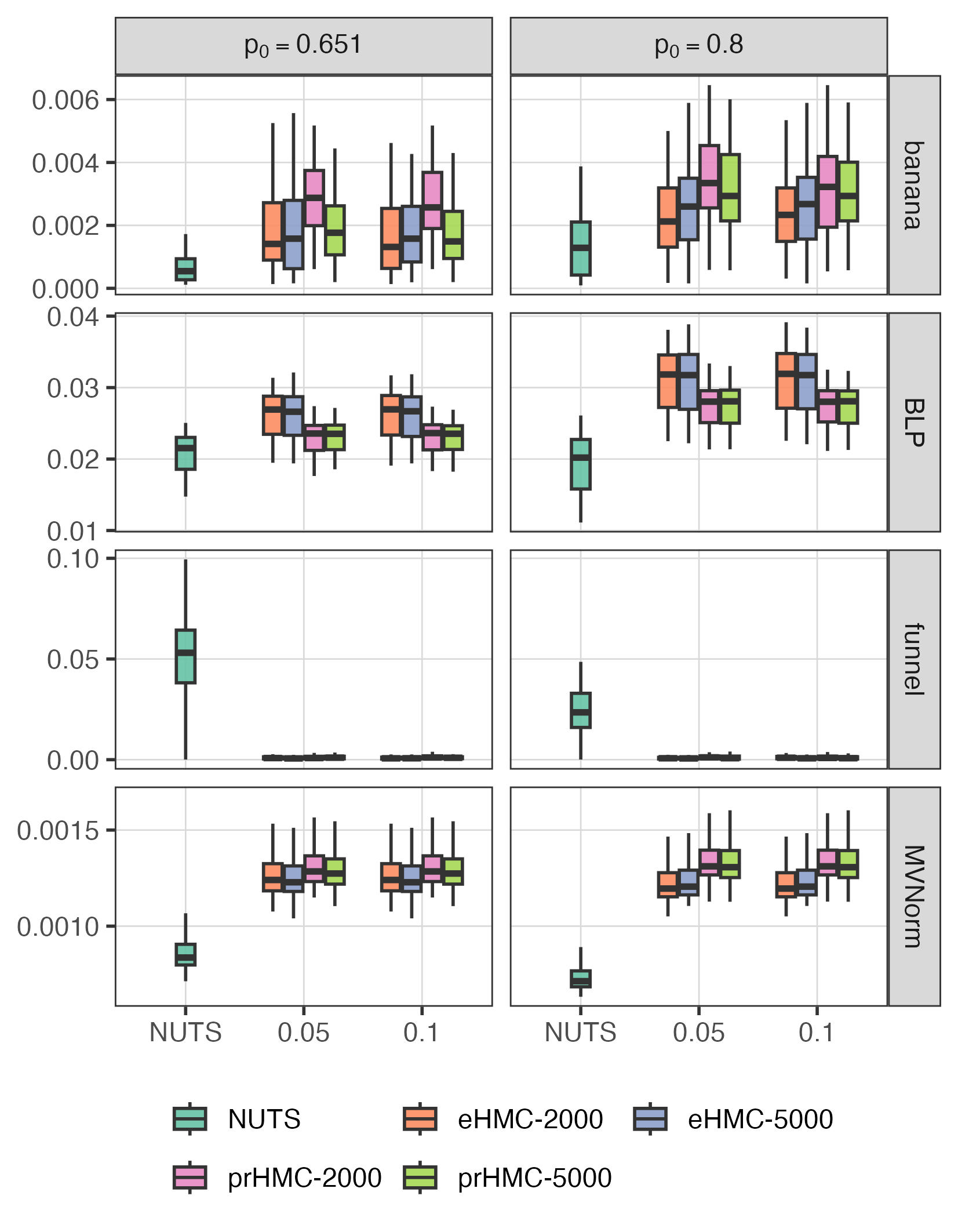}
\includegraphics[width=.49\textwidth]{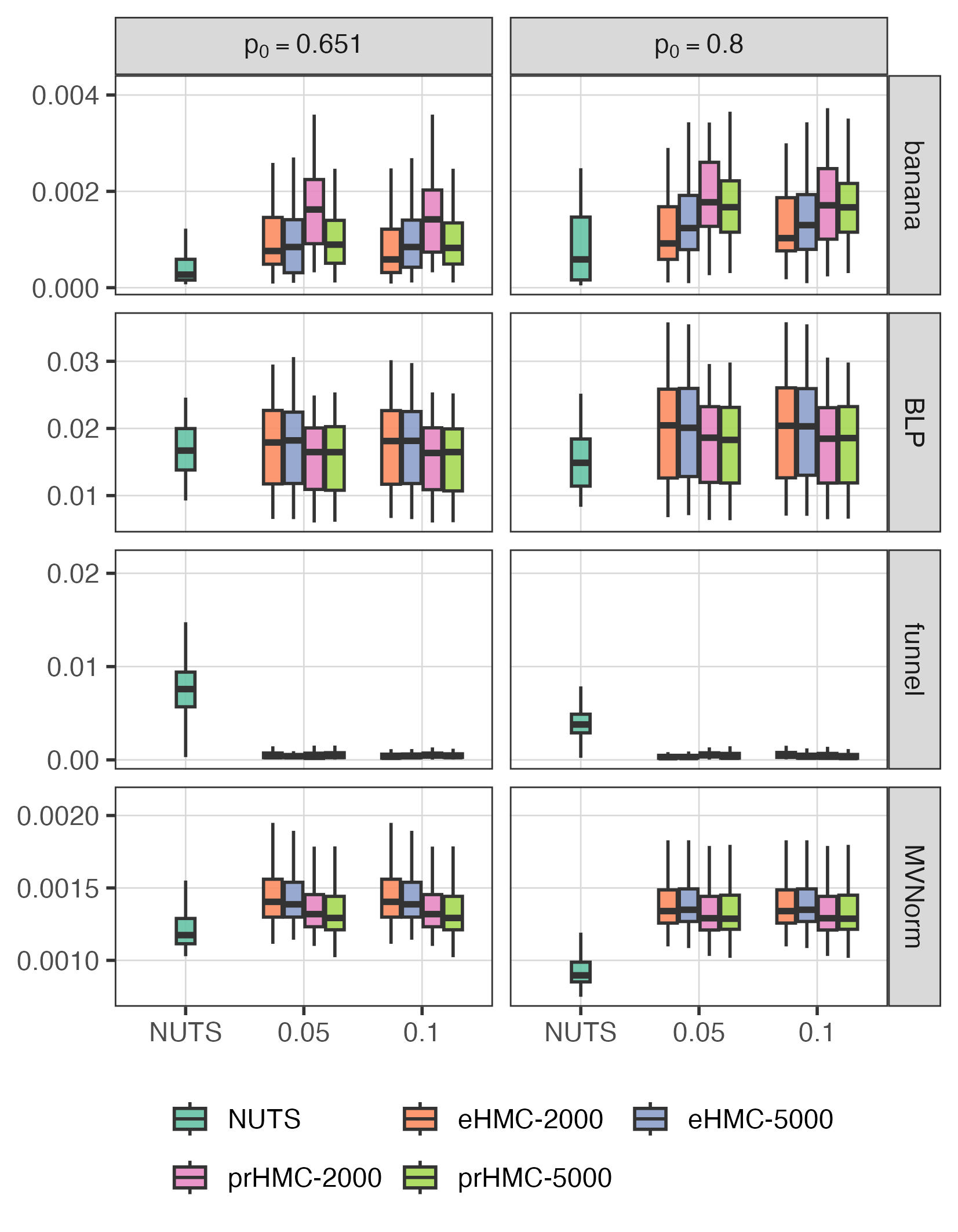}
\caption{Distribution of the Markov chain effective sample size per number of leapfrog steps for mean (left panel) and variance (right panel) estimation, aggregated across dimensions and 20 independent runs of 50 parallel chains of length $T = 2000$. 
prHMC denotes the variant $P_\eta^{(2)}$, where $(v, \varepsilon)$ are partially refreshed with probability $\eta = 0.75$. The same proposal is used for both eHMC and prHMC, trained with $\lambda_0 = 0.8$, $N_0 = 50000$ samples, from which $N \in \{2000, 5000\}$ particles are resampled to construct the empirical distribution on integration parameters. Results are shown for $\Delta/N_0 \in \{5\%, 10\%\}$ in the $x$-axis.
}
\label{fig:prHMCT-ess-norm}
\end{figure}

\begin{figure}[!h]%
\centering
\includegraphics[width=.49\textwidth]{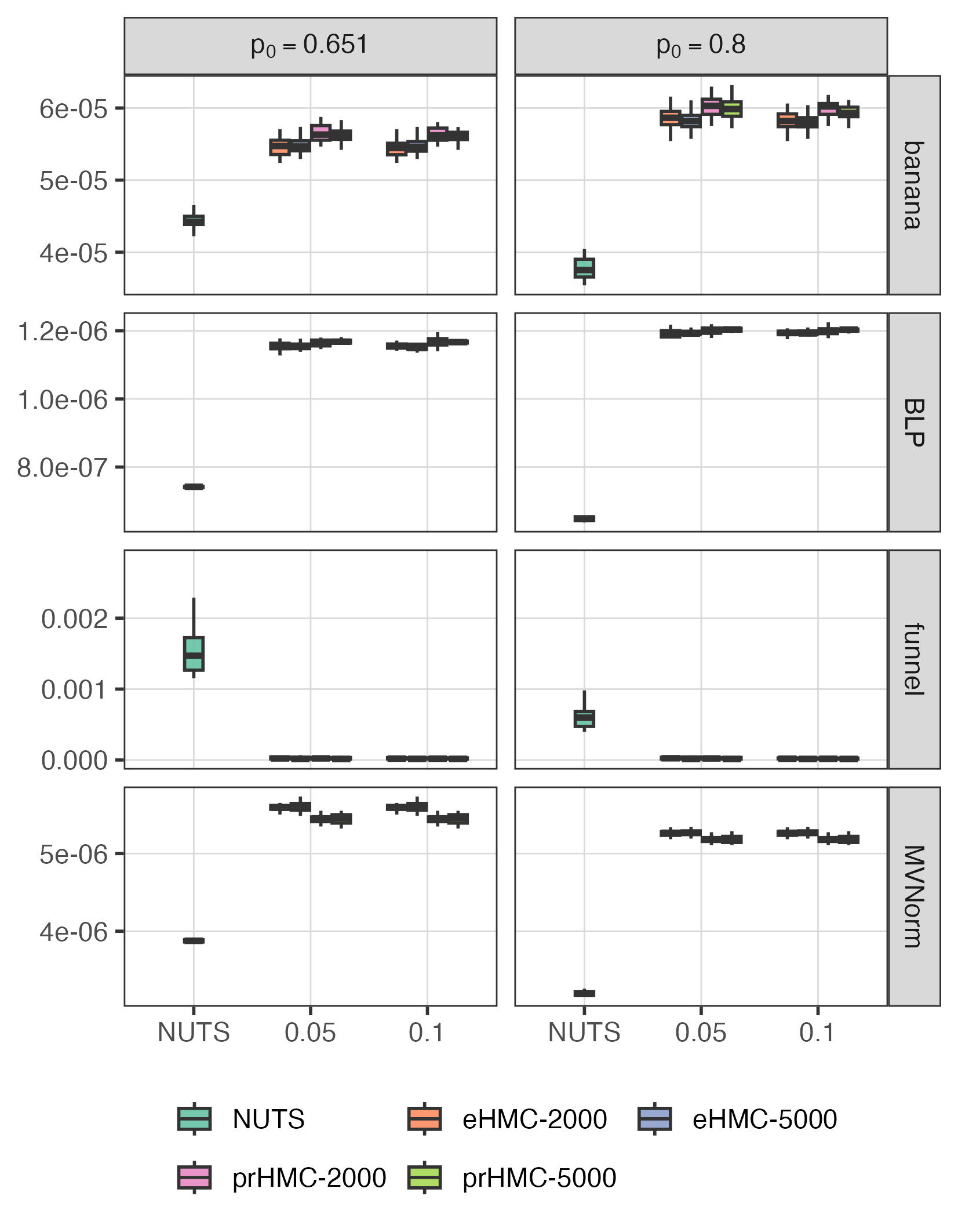}
\includegraphics[width=.49\textwidth]{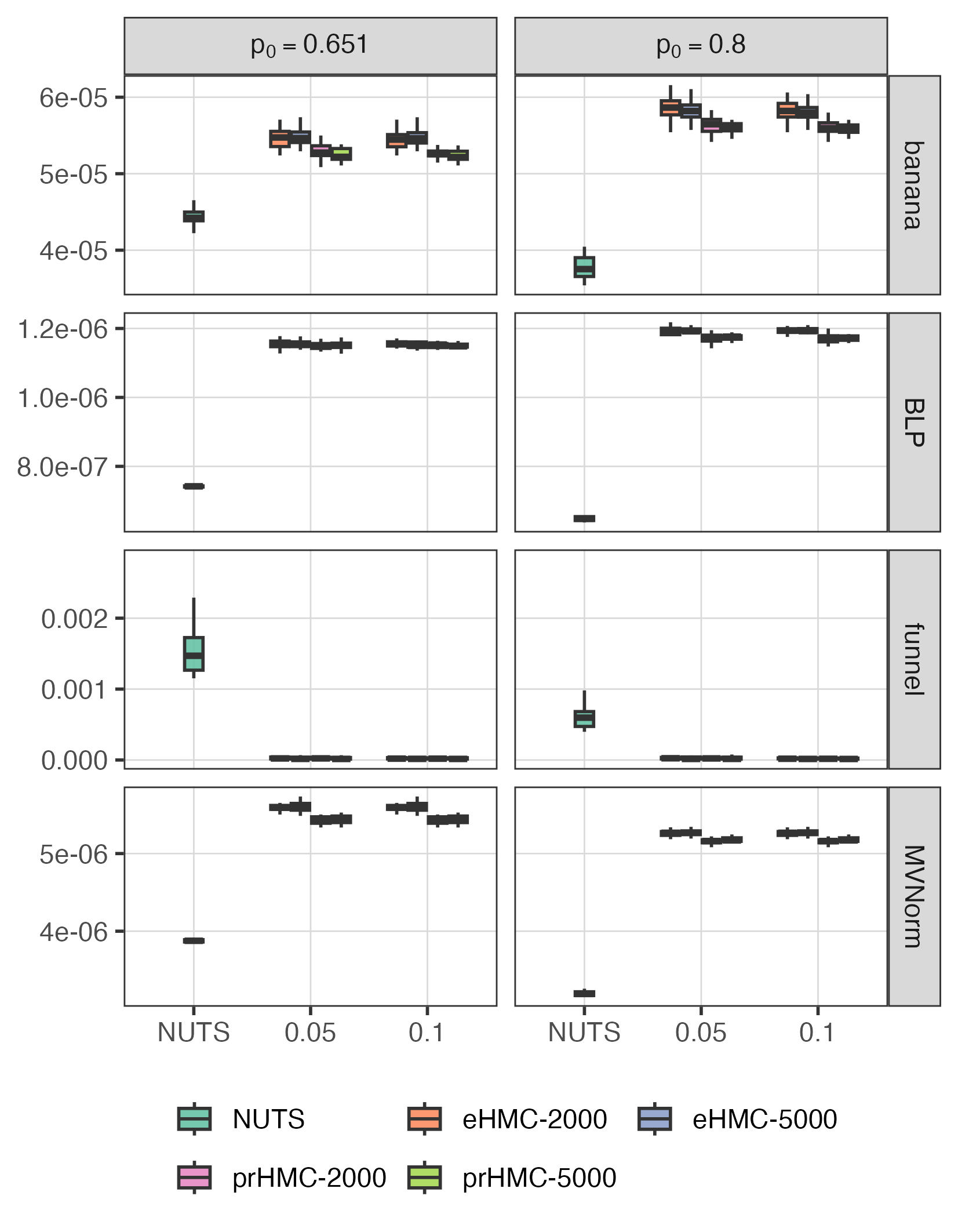}
\caption{Distribution of the expected squared jumping distance over 20 independent runs of 50 parallel chains of length $T = 2000$. The left panel shows the prHMC variant $P_\eta^{(1)}$, where $(v, \varepsilon, L)$ are partially refreshed, and the right panel shows $P_\eta^{(2)}$, where $(v, \varepsilon)$ are partially refreshed, with $\eta = 0.75$. 
Both methods use the same proposal, trained with $\lambda_0 = 0.8$, $N_0 = 50000$ samples, from which $N \in \{2000, 5000\}$ particles are resampled to form the empirical distribution on integration parameters. Results are shown for $\Delta/N_0 \in \{5\%, 10\%\}$ in the $x$-axis.
}
\label{fig:prHMC-esjd-norm}
\end{figure}

\FloatBarrier

\end{appendices}

\end{document}